\documentclass[%
 reprint,
 superscriptaddress,
 amsmath,amssymb,
 aps,
]{revtex4-1}

\usepackage{graphicx}
\usepackage{dcolumn}
\usepackage{bm}
\usepackage{xcolor}
\usepackage{mathrsfs}

\begin{document}


\title{Unifying the landscape of nucleon structure: an infrared-safe evolution scheme
}

\author{Rong Wang}
\email{rwang@impcas.ac.cn}
\affiliation{Institute of Modern Physics, Chinese Academy of Sciences, Lanzhou 730000, China}
\affiliation{School of Nuclear Science and Technology, University of Chinese Academy of Sciences, Beijing 100049, China}

\author{Chengdong Han}
\email{chdhan@impcas.ac.cn}
\affiliation{Institute of Modern Physics, Chinese Academy of Sciences, Lanzhou 730000, China}
\affiliation{School of Nuclear Science and Technology, University of Chinese Academy of Sciences, Beijing 100049, China}

\author{Xurong Chen}
\email{xchen@impcas.ac.cn}
\affiliation{Institute of Modern Physics, Chinese Academy of Sciences, Lanzhou 730000, China}
\affiliation{School of Nuclear Science and Technology, University of Chinese Academy of Sciences, Beijing 100049, China}


\date{\today}

\begin{abstract}
A novel approach for describing the evolution of nucleon structure 
from the low-$Q^2$ regime to the high-$Q^2$ asymptotic region is proposed. 
This infrared-safe scheme modifies the parton distribution evolution 
equations to incorporate the corrections from emergent hadron mass 
mechanisms and parton-parton recombination at low $Q^2$. 
The effective parton mass, generated by dynamical chiral symmetry breaking, 
slows the evolution of parton distributions in the infrared region, 
causing the DGLAP evolution to freeze when $Q^2\ll M_{\rm q/g}$. 
Notably, this scheme renders the high-$Q^2$ parton distributions 
insensitive to the choice of input hadronic scale. 
The parton-parton recombination effect is crucial in suppressing 
the rapid growth of parton distributions at small $x$,
consistent with experimental data. 
When applied to three valence quark distributions derived from a quark model, 
our scheme yields parton distributions that agree well 
with deep-inelastic scattering data in both large-$x$ 
and small-$x$ regions, providing a unified description 
of nucleon structure across the entire $Q^2$ range. 
\end{abstract}

\maketitle


\section{Introduction}
\label{sec:intro}

Carrying the most mass of the visible matter, 
nucleons are the essential building blocks of the universe. 
However, there are still many mysteries about the nucleon, 
such as how do its mass \cite{Wilczek:2012sb,Ji:1994av,Lorce:2017xzd,Metz:2020vxd,Wang:2019mza} 
and spin \cite{Jaffe:1989jz,Ji:1996ek,Ji:2020ena} originate from. 
The nucleon is a composite particle made of quarks and gluons 
under the strong interaction, and the underlying theory of 
the strong interaction is quantum chromodynamics (QCD) 
\cite{Gross:1973id,Politzer:1973fx,Yang:1954ek,Fritzsch:1973pi,tHooft:1972tcz}. 
Due to the confinement mechanism at the long distance, 
solving the QCD problem in the nonperturbative region is notoriously perplexing. 
Nevertheless, the nucleon global properties are believed to be emergent 
from the complex dynamics of quarks and gluons governed by QCD theory 
\cite{Ding:2022ows,Papavassiliou:2022wrb,Binosi:2022djx,Roberts:2021nhw,Carman:2023zke}. 

Measuring the nucleon structure is one practical way 
to understand the emergent phenomenon of the nucleon system. 
Experimentally exploring the internal structure of the nucleon 
in a wide kinematic range is critical for testing the emergent 
mechanism of the nucleon properties in QCD theory. 
To ultimately tackle the nucleon structure and reveal the puzzles 
of the strong interaction, a high-luminosity Electron-Ion Collider
in US (EIC) \cite{AbdulKhalek:2021gbh,Accardi:2012qut} 
is currently under construction. 
An Electron-ion collider in China (EicC) 
\cite{Anderle:2021wcy,Chen:2018wyz,Chen:2020ijn,Wang:2022xad} 
and a plan of JLab 24 GeV upgrade \cite{Accardi:2023chb} also are proposed. 
Undoubtedly, we will acquire a more precise mapping 
of the nucleon structure in the near future. 

The most fundamental quantity for describing the nucleon structure
is parton distribution function (PDF), which gives the probability 
density of the parton (quark or gluon) carrying $x$ fraction of 
the hadron momentum in the infinite momentum frame 
\cite{ParticleDataGroup:2022pth,Bjorken:1968dy,Feynman:1969ej,Bjorken:1969ja}. 
Thanks to the factorization theorem \cite{Collins:1987pm,Collins:1989gx,Sterman:1995fz}, 
the cross section of 
high-energy reaction involving an initial-state hadron 
can be factorized into the product of the parton scattering 
cross section at short distance and the PDF inside the hadron. 
Calculating PDF is a very challenging task, as PDF is of 
nonperturbative origin, related to the long-distance mechanism in a hadron. 
Fortunately, PDF is a universal function, which can be determined 
from the experimental data of many high-energy reactions. 
The lepton deep inelastic scattering (DIS) off the nucleon 
\cite{Taylor:1991ew,Kendall:1991np,Friedman:1991nq} is the 
most prominent way in constraining the nucleon PDFs 
\cite{Pumplin:2002vw,Dulat:2015mca,Ball:2011uy}. 

In theory, it is hard for us to understand why the nucleon PDFs are what they are. 
Nowadays, there are a lot of progresses on this problem 
from nonperturbative QCD methods 
\cite{Yu:2024qsd,Lu:2022cjx,Roberts:2023lap,Chang:2014lva,Chang:2014gga,Ding:2019lwe,Cui:2020tdf,
Ji:2013dva,Ji:2014gla,Ji:2020ect,Ji:2017oey,Ma:2014jla,Ma:2017pxb,Constantinou:2020hdm,Lin:2017snn} 
and many other models \cite{Holt:2010vj,Mineo:2003vc,Kock:2020frx,Wang:2014lua,Steffens:1994hf,Radyushkin:2004mt}. 
Quasi-PDF \cite{Ji:2013dva,Ji:2014gla,Ji:2020ect,Ji:2017oey} 
and pseudo-PDF \cite{Ma:2014jla,Ma:2017pxb} can be evaluated by Lattice QCD (LQCD) simulation, 
however the physics picture is not clear and obvious. 
Usually, it is a much economic way to calculate the PDFs
at low $Q^2$ scale, for there are much fewer components in the nuncleon 
(three-quark and five-quark configurations may just enough). 
Although the factorization formula fails in interpreting the cross section data at low $Q^2$, 
the PDFs exist at low $Q^2$ according to the definition 
of the matrix elements of bilocal light-cone operators. 
The definition of PDF works in the nonperturbative regime 
regardless of how small the $Q^2$ is, 
and we also would like to point that the low-$Q^2$ PDF 
can not be directly connected to the inelastic cross section at low $Q^2$. 

At low $Q^2$, the quark model gives the simplest description of 
the nucleon structure of mere three valence quarks, 
which are the minimal components building a baryon. 
No doubt that it is easier to solve the nucleon structure problem 
at such a low $Q^2$, where the fluctuations of sea quarks 
and gluons are totally frozen out. 
However, to relate the nucleon-structure prediction at low $Q^2$ 
to the experimental measurements, we need an infrared-safe evolution scheme 
for evolving the PDFs from an extremely low $Q^2$ to a hard scale 
that the fatorization formula works. 

The PDF evolution over the $Q^2$ scale is well described with 
Dokshitzer-Gribov-Lipatov-Altarelli-Parisi (DGLAP) equations 
\cite{Dokshitzer:1977sg,Gribov:1972ri,Altarelli:1977zs}. 
The basis of DGLAP equation is the renormalization group equation, 
and the evolution speed is rigorously quantified by the parton splitting vertices in QCD. 
However, the standard DGLAP equations can not be applied at extremely low $Q^2$, 
mainly due to the large strong coupling $\alpha_{\rm s}$ 
and the parton effective mass effect to the splitting kernels. 
We lack of a valid evolution scheme which works from the far-infrared regime 
($Q^2\approx 0$ GeV$^2$) to the asymptotic regime. 

To have a reliable evolution scheme in the infrared region, 
we need to implement the corrections of the nonperturbative QCD effects. 
One outstanding nonperturbative effect is the dynamical chiral symmetry
breaking (DCSB) \cite{Nambu:1961tp,Krein:1990sf,Chang:2011zgy,Roberts:2007ji,Munczek:1994zz,Aguilar:2019uob,Roberts:2021nhw}, 
which is one pillar of the emergent hadron mass mechanism. 
It is one conspicuous feature of QCD, as it provides a mechanism 
of generating mass from nothing. 
The other nonperturbative effect is the effective charge of QCD 
\cite{Deur:2023dzc,Deur:2016tte,Binosi:2016nme,Cui:2019dwv,Cui:2020tdf,Roberts:2021nhw}, 
which is free of Landau pole and saturates in the infrared region. 
The QCD effective charge stops growing as $Q^2$ approaching zero, 
as a result of the mass scale of gluon at low momentum. 
These nonperturbative effects surely alter the DGLAP evolution process 
starting from a very low $Q^2$. 

Another correction at low $Q^2$ to the standard DGLAP evolution is 
the parton-parton recombination correction \cite{Gribov:1983ivg,Mueller:1985wy}. 
In the low-$Q^2$ region, the parton is of large transverse size $\sim 1/Q$, 
and the probability increases for the spatially overlapping of two parton quanta. 
The overlapping probability of two partons is schematically thought 
to be $1/(QR)^2$ ($R$ is average distance between two partons). 
The QCD dynamics of parton-parton recombination processes are quantitatively 
described with the famous Gribov-Levin-Ryskin-Mueller-Qiu-Zhu-Ruan-Shen 
(GLR-MQ-ZRS) correction \cite{Gribov:1983ivg,Mueller:1985wy,Zhu:1998hg,Zhu:1999ht,Zhu:2004xj} 
to the DGLAP equations, 
which include the parton fusion processes. 
The GLR-MQ-ZRS correction is a negative contribution to the fast growing of partons 
thus it appears as a shadowing effect for a photon probe of low $Q^2$ \cite{Mueller:1985wy}. 
Since it is a long evolution distance from $Q^2\sim 0$ GeV$^2$
to a hard scale of experimental measurement, 
the shadowing effect from parton-parton recombination at low $Q^2$ is 
significant in reducing the fast-growing PDFs at small $x$ from splittings. 

In this work, we seek for an infrared-safe evolution scheme from extremely low $Q^2$ 
and test a natural nonperturbative input of three valence quarks from quark model. 
The quark-model description of nucleon structure is revisited in Sec. \ref{sec:quark-model}. 
The standard DGLAP evolution equation is simply reviewed in Sec. \ref{sec:dglap-evolution}. 
Sec. \ref{sec:infrared-safe-evolution} introduces the proposed new evolution scheme 
which works in far-infrared region of $Q^2$ close to zero. 
In Sec. \ref{sec:infrared-safe-evolution}, we discuss in details 
how the nonperturbative effects of DCSB and saturating $\alpha_{\rm s}$ modify 
the DGLAP equations, as well as the parton-parton recombination effect. 
In Sec. \ref{sec:discussion}, we present the evolution result from the suggested 
new evolution scheme with the input PDFs at the infrared scale from quark model. 
The new evolution scheme is found to be a successful bridge connecting 
the nonperturbative regime with the asymptotic regime. 
A summary is provide in Sec. \ref{sec:summary}.

\section{Quark model and origin of parton distributions}
\label{sec:quark-model}

Invented much earlier and as the predecessor of QCD theory, 
the quark model is quite successful in classification of hadrons 
and interpretations of some basic properties of hadrons 
\cite{Gell-Mann:1964ewy,Zweig:1964ruk,Giannini:1990pc,Eichmann:2016yit}.  
The quark model reveals clearly the fundamental flavor symmetry 
and the inner structures of hadrons. 
In terms of the hadron structure, quark model assumes 
that a meson is made of a quark-antiquark pair 
and a baryon is made of three quarks. 
Although it may be just a good approximation, 
the quark model provides the simplest and the most economic 
way in depicting the internal structure of the nucleon. 

At high $Q^2$, as unveiled by high-energy scattering experiments, 
there are plenty of quarks (and antiquarks) and gluons inside the nucleon. 
Where are these partons and the related PDFs originate from? 
PDF is of nonperturbative origin related to 
the long-distance dynamics in the nucleon. 
Given the remarkable success of quark model, 
the quark-model description on the nucleon structure 
serves as the origin of PDFs. 
In the global fit, the initial PDFs at the low scale $Q_0^2$ is called 
the nonperturbative input \cite{Pumplin:2002vw,Dulat:2015mca,Ball:2011uy}, 
which is the input for DGLAP evolution. 
A natural and intuitive nonperturbative input is the three valence 
quark distributions within the quark model 
\cite{Altarelli:1973ff,Parisi:1976fz,Gluck:1977ah,Vainshtein:1976kd,Wang:2014lua}. 
The quark-model picture for the nucleon structure works only 
at a low scale $Q_0^2$ of large resolution, 
where we are confident that only the minimal constituents 
can be resolved in the nucleon. 

The hadronic scale $Q_{\rm h}^2$ of the nucleon is defined 
as the scale at which there are only three valence quarks, 
i.e., at the hadronic scale the nucleon structure is described 
perfectly with the quark model picture. 
The number of sea quarks and gluons increases with the increasing $Q^2$, 
by the reason of the fluctuation processes in QCD theory 
\cite{Altarelli:1977zs,Collins:1988wj}. 
Reversely, at sufficiently low $Q^2$, 
sea quarks and gluons should disappear, 
and there are merely the valence components. 
Quantitatively, where is the hadronic scale? 
In the dynamical parton model by Gl\"{u}ck, Reya and Vogt (GRV), 
all PDFs are radiatively generated from the input of three valence quarks 
and some valence-like gluon distribution 
($\sim 30\%$ momentum fraction) at $Q_0^2=0.26$ GeV$^2$    \cite{Gluck:1998xa}.  
In our previous studies, the principal features of PDFs at high $Q^2$ 
are reproduced from the DGLAP evolution of only 
three valence quark distributions at $Q_0^2\sim 0.07$ GeV$^2$      \cite{Wang:2016sfq}.  
According to the definition and above discussions, 
the hadronic scale is necessarily in the far-infrared region 
($Q_{\rm h}^2 < 0.1$ GeV$^2$). 

In order to test the idea that the nonperturbative input of quark model is 
the origin of PDFs, an infrared-safe evolution scheme is quite demanding, 
for the hadronic scale is in the infrared domain, 
where the quark-model picture for the hadron is valid.

\section{DGLAP evolution equations}
\label{sec:dglap-evolution}

The $Q^2$-dependence of PDFs is successfully described with DGLAP equations 
\cite{Dokshitzer:1977sg,Gribov:1972ri,Altarelli:1977zs}. 
In history, the variation of deep inelastic structure function 
over the $Q^2$ scale is firstly explained in perturbation theory 
in the principal logarithmic approximation \cite{Gribov:1972ri}.
With the advent of QCD, the evolution of structure function 
over $Q^2$ is rigorously formulated with the operator-product expansion
in the language of renormalization group equations. 
A much clear picture is provided by Altarelli and Parisi, 
with the parton density evolution over $Q^2$ 
deduced from the basic splitting vertices of QCD \cite{Altarelli:1977zs}. 
With the increasing resolution power of virtual photon probe of $Q^2$, 
one could resolve more partons of smaller size from the fluctuation 
of splitting processes. In short, 
the variation of parton density at different scales is governed 
by the strong interactions among quarks and gluons. 
Up to now, DGLAP equation has been one of the most 
successful tests of QCD theory at high $Q^2$. 

The DGLAP equations for $Q^2$-dependence of PDFs are 
a set of integro-differential equations. 
It is usually convenient to write the master equations 
in terms of non-singlet ($q_i^{\rm NS}=q_i-\bar{q}_i$), 
singlet ($q^{\rm S}=\sum_i(q_i+\bar{q}_i)$), and gluon distributions. 
The DGLAP evolution equations are written as, 
\begin{equation}
\begin{split}
   \frac{dq_{\rm i}^{\rm NS}}{d{\rm ln}(Q^2)} =
   \frac{\alpha_{\rm s}(Q^2)}{2\pi}
   P_{\rm qq} \otimes q_{\rm i}^{\rm NS},
\end{split}
\label{eq:AP-NS}
\end{equation}
for non-single quark distributions, and 
\begin{equation}
\begin{split}
   \frac{d}{d{\rm ln}(Q^2)} \left(
      \begin{array}{c}
         q^{\rm S} \\
         g \\
      \end{array}
   \right)
   = \frac{\alpha_{\rm s}(Q^2)}{2\pi} \left(
      \begin{array}{cc}
         P_{\rm qq} & 2n_{\rm f}P_{\rm qg} \\
         P_{\rm gq} & P_{\rm gg} \\
      \end{array}
   \right) \otimes \left(
      \begin{array}{c}
         q^{\rm S} \\
         g \\
      \end{array}
   \right),
\end{split}
\label{eq:AP-singlet-gluon}
\end{equation}
for singlet quark and gluon distributions, 
where $P_{\rm qq}$, $P_{\rm qg}$, $P_{\rm gq}$ and 
$P_{\rm gg}$ are the standard DGLAP splitting kernels, 
and $n_{\rm f}$ denotes the number of active quark flavors in the evolution.

\section{An infrared-safe evolution scheme}
\label{sec:infrared-safe-evolution}

In the classic DGLAP equations discussed above, 
the nonperturbative effects are not considered at all, 
thus it can not be applied at low energy scale. 
In this section, we look for a solid evolution scheme 
which works in the whole $Q^2$ range, 
even down to around 0 GeV$^2$. 

The QCD theory exhibits three prominent features: 
asymptotic freedom, approximate chiral symmetry and confinement. 
In the nonperturbative region, the chiral symmetry is dynamically broken, 
and the quarks and gluons are confined in the hadron of the size near 1 fm. 
DCSB arises from the nonperturbative mechanism of QCD, 
and it dresses up the parton with large effective mass 
\cite{Krein:1990sf,Chang:2011zgy,Roberts:2007ji,Munczek:1994zz,Aguilar:2019uob,Roberts:2021nhw}. 
The parton effective mass eventually influences 
the evolution kernels of parton radiations. 
Owing to DCSB, the strong coupling constant $\alpha_{\rm s}$ is finite 
and free of Landau pole in the infrared region, 
according to a definition of effective charge in QCD 
\cite{Deur:2023dzc,Binosi:2016nme,Roberts:2021nhw}. 
Actually, the $\alpha_{\rm s}$ saturates at $k^2=0$ GeV$^2$ 
under the nonperturbative mechanism solved with Dyson-Schwinger equations. 
To extend DGLAP equations to the infrared regime, 
the discussed nonperturbative QCD effects should be taken into account. 

Still, due to the large parton size at small momentum $Q^2$      \cite{Mueller:1985wy}, 
the nonlinear effect from parton overlapping 
should also be considered for the new evolution scheme 
that is extended to the infrared region.

\subsection{Dynamical chiral symmetry breaking}
\label{subsec:DCSB}

\begin{figure}[htbp]
\begin{center}
\includegraphics[width=0.45\textwidth]{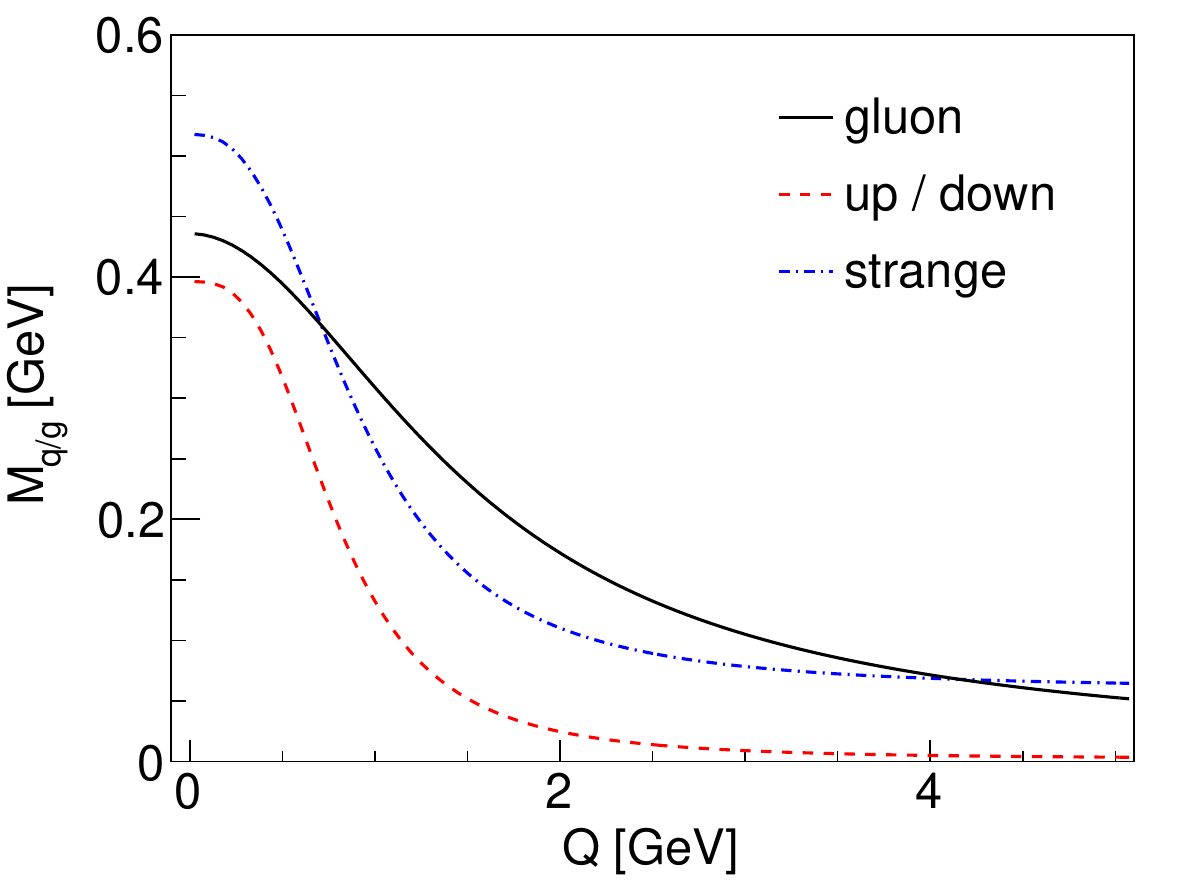}
\caption{
  The mass functions of dressed quark and gluon as a function
  of momentum $Q$ from Dyson-Schwinger equations \cite{Aguilar:2019uob,Roberts:2021nhw}.
}
\label{fig:mass-functions}
\end{center}
\end{figure} 

Here we discuss how the effective parton mass modify the DGLAP 
evolution kernels. Let us take a closer look at the mass functions 
of partons from nonperturbative QCD methods. 

DCSB is clearly revealed in the two-point Green function 
from both the Continuum Schwinger-functional Methods 
(CSMs) and the LQCD approach \cite{Chang:2011zgy}, and the amazing consistence 
is found between these nonperturbative methods. 
The simple picture is that the massless parton obtains a 
large mass through the strong interactions with its own gluon field, 
and DCSB is vividly shown in the mass functions of quark and gluon   \cite{Aguilar:2019uob,Roberts:2021nhw}. 
In CSMs, the dressed gluon or quark propagator is a direct 
solution of the gap equations. 
Fig. \ref{fig:mass-functions} shows the dressed parton mass functions 
as a function of the momentum, which are taken from the results of CSMs. 
One sees a smooth transition from nearly zero mass at high momentum
to a saturating mass around hundreds of MeV at low momentum. 
The saturating mass of quark is simply around the constituent quark mass. 

In this work, the modification of DGLAP evolution due to the parton 
effective mass at low momentum is considered. 
The DGLAP splitting function of quark in the infinite momentum frame is written as, 
\begin{equation}
\begin{split}
    &\frac{E_k}{E_l}|M_{l\rightarrow kl^{\prime}}|^2
    \left[\frac{1}{E_l-E_k-E_{l^{\prime}}}\right]^2
    \left[\frac{1}{2E_k}\right]^2
    \frac{d^3l^{\prime}}{(2\pi)^32E_{l^{\prime}}} \\
    &= P_{qq}\frac{dk_T^2}{k_T^2},
\end{split}
\label{eq:splitting-kernel}
\end{equation}
for the process $q(l) \rightarrow q(k) + g(l^{\prime})$.
Following Zhu and Wang's idea \cite{Zhu:2019ydy}, let us look at the $k_T$-dependent part 
of the splitting kernel, the modified massive propagator is given by, 
\begin{equation}
\begin{split}
    \frac{dk_T^2}{k_T^2+M_q^2}=\frac{d(k_T^2+M_q^2)}{(k_T^2+M_q^2)}=\frac{d\tilde{k}_T^2}{\tilde{k}_T^2},
\end{split}
\label{eq:massive-propagator}
\end{equation}
where $M_q^2$ is from the energy expression of the parton.  
From the renormalization group equation, 
$k_T^2$ of the parton is replaced by the renormalization scale $\mu^2=Q^2$. 
Thus, considering the massive parton, 
the evolution scale that DGLAP equations depend on should be 
the new scale $\tilde{Q}^2$, which is related to $Q^2$ by,  
\begin{equation}
\begin{split}
    \tilde{Q}^2 = Q^2 + M_q^2.
\end{split}
\label{eq:new-sclae}
\end{equation}

Therefore, to implement the parton effective mass effect from DCSB, 
one simply takes a new evolution path (${\rm ln}\tilde{Q}^2$) for the evolution, 
which is written as, 
\begin{equation}
\begin{split}
    \frac{dxf}{dln\tilde{Q}^2} = \frac{\alpha}{2\pi} P_{ff} \otimes xf. \\
\end{split}
\label{eq:evolution-mass-effect-1}
\end{equation}
With Eq. (\ref{eq:new-sclae}) and writing the equation in terms of 
the factorization scale $Q^2$ in experiment, 
one gets the modified evolution equation for the massive parton, 
which is written as, 
\begin{equation}
\begin{split}
    \frac{dxf}{dlnQ^2} = \frac{Q^2}{Q^2+M_q^2}
    \times \frac{\alpha}{2\pi} P_{ff} \otimes xf.
\end{split}
\label{eq:evolution-mass-effect-2}
\end{equation} 
One finds that the factor $Q^2/(Q^2+M_q^2)$ in the equation slows down
the evolution process when $Q^2$ is a small value, 
while it is unity when $Q^2$ is a large value ($Q^2\gg M_q^2$).

\subsection{Effective charge of QCD}
\label{subsec:sat-alphas}

The running strong coupling is an essential element for evolution equations. 
Due to the divergent Landau pole, any calculations in the far 
infrared region based on perturbative strong coupling can not be made. 
To solve this problem, we capitalize on the existence of 
a renormalization-group-invariant (RGI) and process-independent (PI) 
effective charge in QCD \cite{Binosi:2016nme}. 
It is obtained from a self-consistent solution of Dyson-Schwinger Equations (DSEs) 
and the gluon propagator from LQCD \cite{Cui:2019dwv}. 
The QCD effective charge and DCSB are two nonperturbative phenomena 
which interplay with each other. 
The Landau pole is screened in QCD by the dynamical generation of a gluon mass. 
The effective charge saturates to a finite value at infrared momentum, 
providing an infrared completion of strong coupling $\alpha_{\rm s}$. 
Moreover, the effective charge smoothly transit to 
the perturbative QCD coupling in the ultraviolet region. 

For our proposed infrared-safe evolution scheme, 
we take the effective charge tuned with the up-to-date 
DSEs and LQCD calculations. 
According to QCD effective charge, a parametrization of 
saturating $\alpha_{\rm s}$ is given by  
\cite{Binosi:2016nme,Cui:2019dwv,Cui:2020tdf,Roberts:2021nhw},  
\begin{equation}
\begin{split}
   \alpha_{\rm s} = \frac{\gamma_m\pi}{{\rm ln}[\mathscr{K}^2(k^2)/\Lambda_{\rm QCD}^2]}, \\
   \mathscr{K}(y) = \frac{a_0^2 + a_1y + y^2}{b_0+y}, \\
\end{split}
\label{eq:effective-charge}
\end{equation}
in which $\gamma_m=4/\beta_0$, $\beta_0=11-(2/3)n_f$, and $\Lambda_{QCD}=0.234$.
This saturating strong coupling represents the interaction strength in QCD 
at any given momentum scale. 

\begin{figure}[htbp]
\begin{center}
\includegraphics[width=0.45\textwidth]{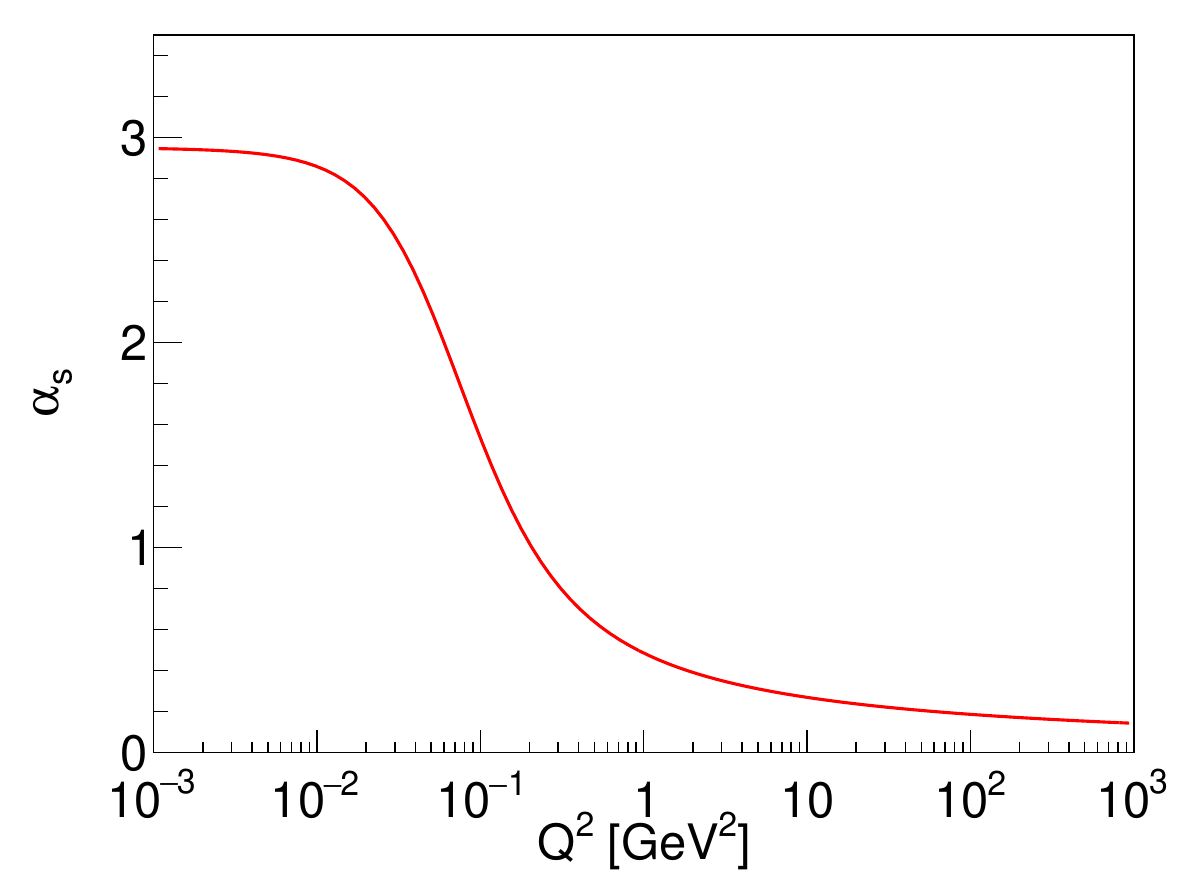}
\caption{
  The saturated strong coupling (in infrared limit) from Dyson-Schwinger equations
  \cite{Binosi:2016nme,Cui:2019dwv,Cui:2020tdf,Roberts:2021nhw}.
}
\label{fig:alphas}
\end{center}
\end{figure}

\subsection{Parton-parton recombination effect}
\label{subsec:recombination-effect}

The standard DGLAP evolution leads to a rapid growth of 
gluon distribution at small $x$, 
which eventually breaks up the cross section unitarity \cite{Froissart:1961ux,Martin:1962rt}.   
For parton distribution evolution at small $x$, 
especially starting from a low scale $Q_0^2$, 
the parton-parton recombination correction is vital. 
A simple criteria is that when $xg\frac{1}{Q^2}>R_{\rm p}^2$, 
the gluons within a unit of rapidity begins 
to spatially overlap in the thin disc of the proton \cite{Mueller:1985wy} . 
One expects the fusions or recombinations of partons to happen at small-$x$. 
The parton-recombination correction to DGALP equations was initiated 
by Gribov, Levin and Ryskin \cite{Gribov:1983ivg}, and latter quantitatively evaluated 
by Mueller and Qiu in covariant field theory \cite{Mueller:1985wy}. 
All kinds of parton recombinations (gluon-gluon, quark-gluon, and quark-quark) are then 
derived in time-ordered perturbation theory \cite{Zhu:1998hg,Zhu:1999ht,Zhu:2004xj} . 
Parton-parton recombination is a high-twist effect proportional to $1/Q^2$, 
which is also taken into account for our infrared-safe evolution scheme. 

Combine all the effects from DCSB, QCD effective charge and 
parton-parton recombination, the modified DGLAP equations 
for the infrared-safe evolution scheme are given by, 
\begin{equation}
\begin{split}
   \frac{dxq_{\rm i}^{\rm NS}}{d{\rm ln}(Q^2)} =
   \frac{Q^2}{Q^2+M_{\rm q}^2}\frac{\alpha_{\rm s}(Q^2)}{2\pi}
   P_{\rm qq} \otimes xq_{\rm i}^{\rm NS},
\end{split}
\label{eq:modifiedAP-NS}
\end{equation}
for the flavor non-singlet quark distributions, 
\begin{equation}
\begin{split}
   \frac{dx\bar{q}_{\rm i}}{d{\rm ln}(Q^2)} =
   \frac{Q^2}{Q^2+M_{\rm q}^2}\frac{\alpha_{\rm s}(Q^2)}{2\pi}
   \left[
   P_{\rm qq} \otimes x\bar{q}_{\rm i} + P_{\rm qg} \otimes xg
   \right] \\
   -\frac{Q^2}{Q^2+M_{\rm q}^2}\frac{\alpha_{\rm s}^2(Q^2)}{4\pi R^2Q^2}\int_{x}^{1/2}
   \frac{dy}{y}xP_{\rm gg\rightarrow\bar{q}}(x,y)[yg(y,Q^2)]^2 \\
   +\frac{Q^2}{Q^2+M_{\rm q}^2}\frac{\alpha_{\rm s}^2(Q^2)}{4\pi R^2Q^2}\int_{x/2}^{x}
   \frac{dy}{y}xP_{\rm gg\rightarrow\bar{q}}(x,y)[yg(y,Q^2)]^2,
\end{split}
\label{eq:modifiedAP-sea}
\end{equation}
for the sea quark distributions, and 
\begin{equation}
\begin{split}
   \frac{dx\bar{q}_{\rm i}}{d{\rm ln}(Q^2)} =
   \frac{Q^2}{Q^2+M_{\rm g}^2}\frac{\alpha_{\rm s}(Q^2)}{2\pi}
   \left[
   P_{\rm gq} \otimes x\Sigma + P_{\rm gg} \otimes xg
   \right] \\
   -\frac{Q^2}{Q^2+M_{\rm g}^2}\frac{\alpha_{\rm s}^2(Q^2)}{4\pi R^2Q^2}\int_{x}^{1/2}
   \frac{dy}{y}xP_{\rm gg\rightarrow g}(x,y)[yg(y,Q^2)]^2 \\
   +\frac{Q^2}{Q^2+M_{\rm g}^2}\frac{\alpha_{\rm s}^2(Q^2)}{4\pi R^2Q^2}\int_{x/2}^{x}
   \frac{dy}{y}xP_{\rm gg\rightarrow g}(x,y)[yg(y,Q^2)]^2,
\end{split}
\label{eq:modifiedAP-g}
\end{equation}
for the gluon distribution, 
where $P_{\rm qq}$, $P_{\rm qg}$, $P_{\rm gg}$, $P_{\rm gq}$ are 
the DGLAP splitting kernels, and $P_{\rm gg\rightarrow\bar{q}}$, 
$P_{\rm gg\rightarrow\bar{g}}$ are the gluon-gluon recombination coefficients. 
The factor $1/(4\pi R^2)$ in the equations is for two-parton density normalization, 
and $R$ is the correlation length of two interacting partons.  
$R$ is supposed to be smaller than the proton radius. 
In a previous analysis, $R$ is determined to be 3.98 GeV$^{-1}$      \cite{Wang:2016sfq}. 
Note that the integral terms $\int_x^{1/2}$ in above equations 
disappear when $x>1/2$. $\Sigma$ in Eq. (\ref{eq:modifiedAP-g}) denotes 
$\Sigma(x,Q^2)\equiv \Sigma_i [q_i(x,Q^2) + \bar{q}_i(x,Q^2)]$. 

The factor $Q^2/(Q^2+M_{\rm q/g}^2)$ in the modified DGLAP equations 
arises from the massive parton from DCSB, and it quenches the evolution at low $Q^2$. 
The saturating $\alpha_{\rm s}$ extends the QCD interaction strength 
to infrared region, which is crucial for the suggested infrared-safe evolution scheme. 
For simplicity of calculation, we take only the gluon-gluon recombination correction 
for the evolution equations. 
As shown in our previous study, ignoring the quark-quark and quark-gluon
recombination processes does not change much the prediction ($<5\%$) \cite{Chen:2014nba},  
for the gluon distribution is definitely dominant at small $x$.

\subsection{A natural choice of hadronic scale}
\label{subsec:hadronic-scale}

\begin{figure}[htbp]
\begin{center}
\includegraphics[width=0.45\textwidth]{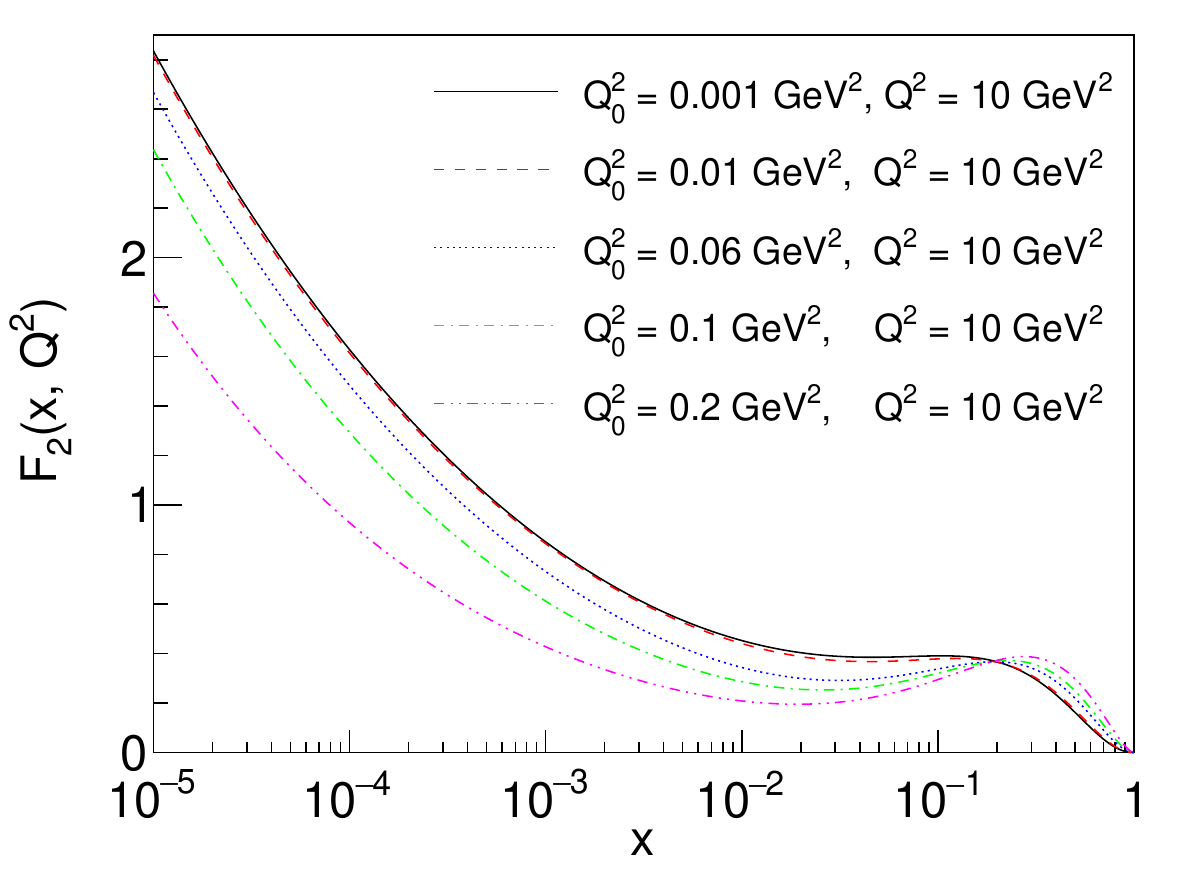}
\caption{
  The predicted structure function $F_2$ at $Q^2=10$ GeV$^2$,
  with different input scale $Q_0^2$ (the hadronic scale).
  The infrared-safe evolution scheme is used for the evolution calculations.
}
\label{fig:F2_with_diff_Q0}
\end{center}
\end{figure}

The hadronic scale is defined as the scale at which 
the nucleon is described well with mere three valence quarks 
as implied by the quark model. 
It is another important parameter for the nonperturbative input of three valence quarks. 
Although the definition is clear, 
the value of the hadronic scale is quite uncertain so far. 
Different groups use the different hadronic scales. 

Fortunately, with the suggested new evolution scheme, 
the evolved PDFs at high $Q^2$ do not change much 
with the hadronic scale, once the hadronic scale is under $M_{\rm q}^2$, 
thanks to the freeze of the evolution at low $Q^2$ in our proposed scheme. 
Fig. \ref{fig:F2_with_diff_Q0} shows the structure function at 
$Q^2=10$ GeV$^2$ with the same nonperturbative input 
but very different hadronic scales. 
One sees that the evolution results are almost identical 
with the hadronic scale set at $0.01$ GeV$^2$ and $0.001$ GeV$^2$. 
The evolution of PDFs is almost completely frozen 
when $Q^2$ is smaller than $0.01$ GeV$^2$. 

The freeze of the evolution at low $Q^2$ is not difficult to understand. 
In the proposed scheme, the evolution path is ${\ln}\tilde{Q}^2$. 
Thus the evolution distance from $Q_0^2$ to $Q_1^2$ is 
$L\equiv\int_{Q^2=Q_0^2}^{Q^2=Q_1^2}  d{\rm ln}\tilde{Q}^2 
= {\rm ln}(Q_1^2+M_{\rm q}^2) - {\rm ln}(Q_0^2+M_{\rm q}^2)$, 
where $M_{\rm q}$ denotes the parton effective mass. 
It is quite clear that the evolution distance $L$ stop increasing 
as $Q_0^2$ approaches zero. 

With the suggested infrared-safe evolution scheme, 
one can simply choose a natural hadronic scale of $Q_0^2=0$ GeV$^2$. 
With this natural hadronic scale, 
the evolution distance to $Q_1^2$ is frozen at 
$L={\rm ln}(Q_1^2+M_{\rm q}^2) - {\rm ln}(M_{\rm q}^2)$. 
We argue that at $Q_0^2=0$ GeV$^2$, 
it is very safe to regard the nucleon as an object containing 
just the minimum components (three quarks) required for a color-singlet. 
We take the natural hadronic scale for the new evolution scheme 
in the following calculations.

\section{Results and discussions}
\label{sec:discussion}

\begin{table}[h]
    \caption{The settings of three evolution schemes performed in this work.
             The Scheme-C is the proposed infrared-safe evolution scheme
             with full nonperturbative and parton-recombination corrections.}
        \renewcommand\arraystretch{1.5}
		\begin{tabular}{ccccc}
        \hline\hline
          Scheme  & $Q_0^2$ [GeV$^2$] & $M_{\rm q/g}$  &  $\alpha_s$    &  Parton fusion \\
        \hline
           A      &    0.06         &   Massless   &   Perturbative &    W/O   \\
           B      &    0.06         &    Dressed   &   Saturating   &     W/O   \\
           C      &   0.001         &    Dressed   &   Saturating   &     W   \\
        \hline\hline
		\end{tabular}
    \label{tab:evolution-schemes}
\end{table}

To understand how the nonperturbative and parton-recombination effects 
influence the DGLAP evolution, 
we have performed the standard DGLAP evolution and 
the evolution's with different corrections switched on. 
The settings of three different evolution schemes are 
listed in Table \ref{tab:evolution-schemes}, 
respectively named scheme-A, B and C for the convenient of discussions. 
Scheme-A is the standard DGLAP evolution; 
Scheme-B is the evolution with the massive parton correction 
and the saturating $\alpha_{\rm s}$; 
Scheme-C is the evolution with the massive parton correction, 
the saturating $\alpha_{\rm s}$ and the parton-parton recombination correction, 
which is the suggested infrared-safe evolution scheme given in this work. 

We take the same nonperturbative input for the three different evolution schemes. 
For simplicity, we use a simple input of three valence quark distributions 
from the maximum-entropy method (MEM) \cite{Wang:2014lua}, since we focus mainly 
on the extension of the evolution equation to low-$Q^2$ domain.  
The MEM nonperturbative input is inferred 
from the constraints of quark model and confinement, 
which is found to be consistent with the experimental measurements 
and the global fits of PDFs \cite{Wang:2014lua}. 

Now let us choose a proper hadronic scale for each evolution scheme. 
The hadronic scale is defined as at the scale where the nucleon consists 
only three valence quarks, i.e., 
the second moment $\left<\Sigma_i xV_i\right>$ of valence quarks is one at the hadronic scale. 
The evolution of the moment is given by the renormalization group equation \cite{Altarelli:1977zs}. 
Hence once we know the valence moment at a high $Q^2$, 
we can get the hadronic scale $Q_0^2$ for the standard DGLAP evolution 
by requiring $\left<xu_V+xd_V\right>(Q_0^2)=1$. 
From CT14(LO) PDFs of a global fit \cite{Dulat:2015mca}, 
the second moment of valence distributions is 0.37 at $Q^2=4$ GeV$^2$. 
If we take the perturbative $\alpha_s$ \cite{Gluck:1998xa} for the moment evolution, 
the hadronic scale is found to be 0.064 GeV$^2$, 
and if we take the saturating $\alpha_s$  \cite{Binosi:2016nme,Cui:2019dwv,Cui:2020tdf,Roberts:2021nhw},   
the hadronic scale is found to be 0.059 GeV$^2$. 
Therefore, we choose the hadronic scale $Q_0^2=0.06$ GeV$^2$  
for scheme-A of the standard DGLAP evolution. 
For scheme-B, the hadronic scale is also at 0.06 GeV$^2$, around $M_{\rm q}^2$. 
For scheme-C, the suggested infrared-safe evolution scheme, 
we choose the natural hadronic scale $Q_0^2=0$ GeV$^2$. 
For the practical computations, the hadronic scale is set at 0.001 GeV$^2$. 
As shown in Fig. \ref{fig:F2_with_diff_Q0}, 
the PDF evolution in scheme-C almost stops under 0.01 GeV$^2$.

\begin{figure}[htbp]
\begin{center}
\includegraphics[width=0.45\textwidth]{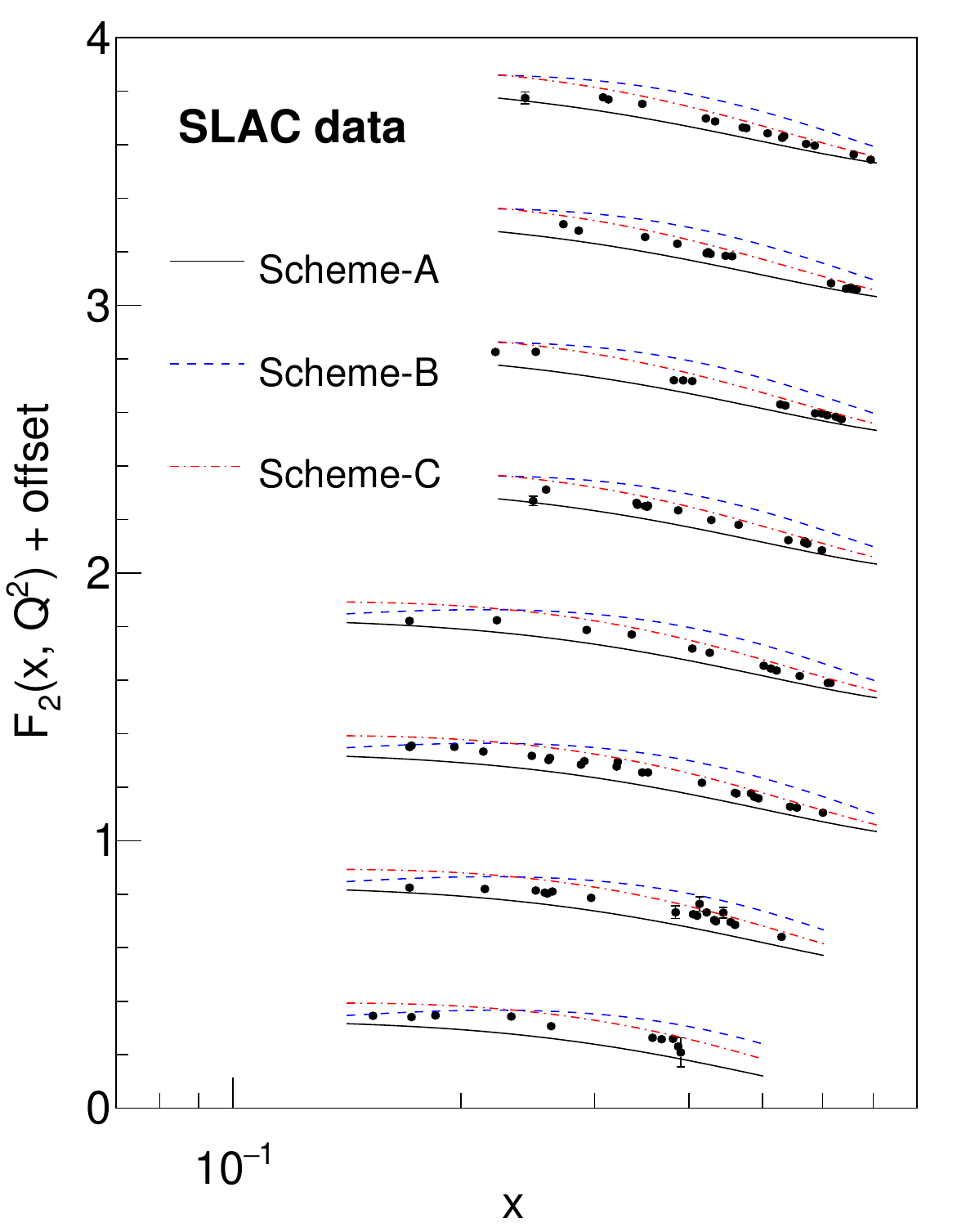}
\caption{
  The predicted structure function $F_2$ in the large-$x$ region
  at high $Q^2$, compared to SLAC data \cite{Whitlow:1991uw}.
  From bottom to top, the $Q^2$ scales of the data are at 4.28, 
  4.71, 5.21, 5.61, 5.98, 6.28, 6.7, and 7.21 GeV$^2$, respectively.
  The solid, dashed and dashed-dotted curves show the results 
  from Scheme-A, Scheme-B and Scheme-C, respectively. 
  See the main text and Table \ref{tab:evolution-schemes} for more explanations. 
  }
\label{fig:F2_at_largex}
\end{center}
\end{figure}

\begin{figure}[htbp]
\begin{center}
\includegraphics[width=0.45\textwidth]{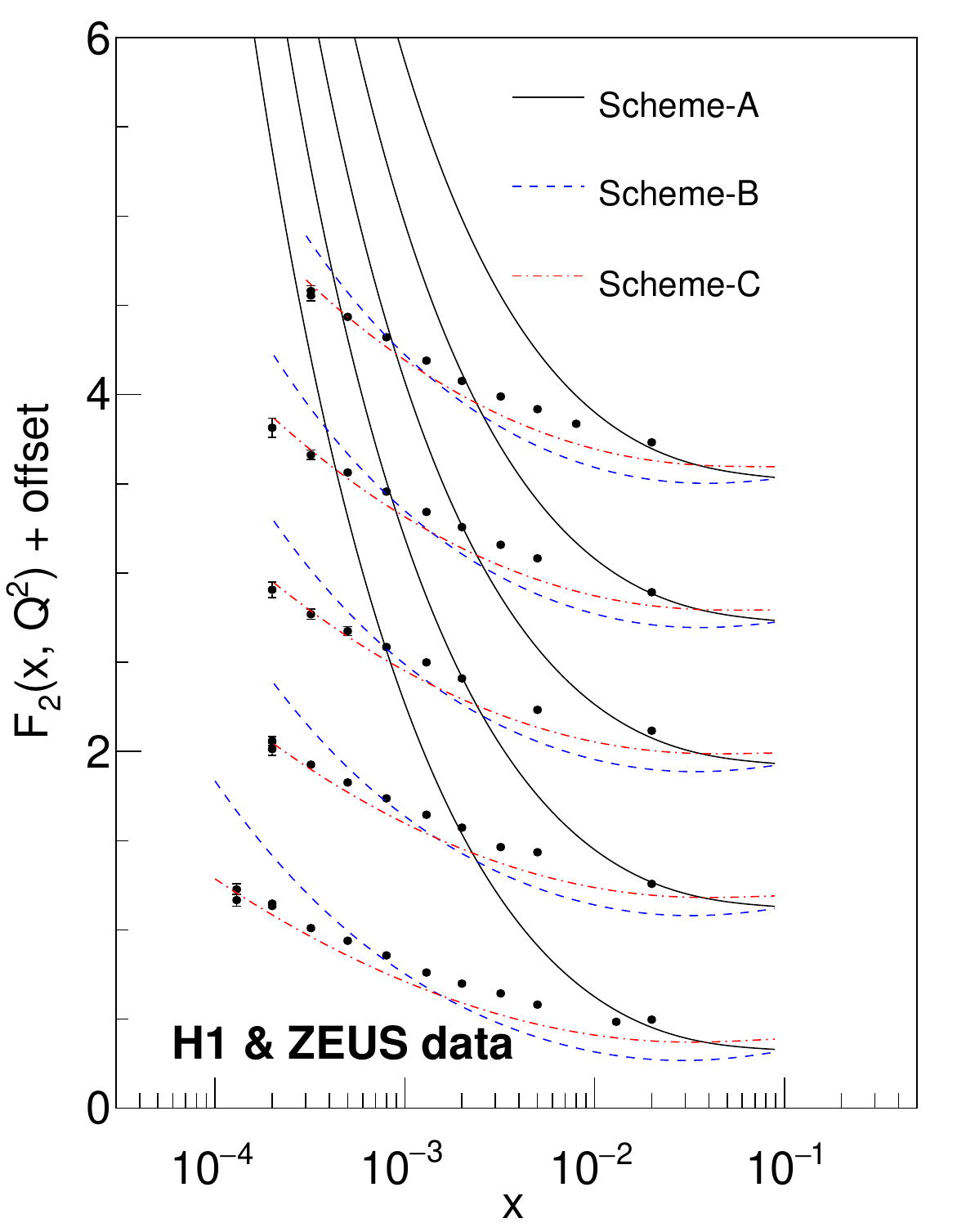}
\caption{
  The predicted structure function $F_2$ in the small-$x$ region
  at high $Q^2$, compared to H1 and ZEUS data \cite{H1:2009pze}.
  From bottom to top, the $Q^2$ scales of the data are at 6.5, 8.5, 
  10, 12, and 15 GeV$^2$, respectively.  
  The solid, dashed and dashed-dotted curves show the results 
  from Scheme-A, Scheme-B and Scheme-C, respectively. 
  See the main text and Table \ref{tab:evolution-schemes} for more explanations. 
}
\label{fig:F2_at_smallx}
\end{center}
\end{figure}

Scheme-A is the standard DGLAP evolution for a reference, 
which takes no account of the nonperturbative effects 
(DCSB and saturating $\alpha_{\rm s}$) 
and the nonlinear correction from parton-parton recombination. 
The PDFs are directly connected with the structure function $F_2$, 
which can be measured in the DIS experiments. 
In Figs. \ref{fig:F2_at_largex} and \ref{fig:F2_at_smallx}, 
we show the predicted $F_2$ at high $Q^2$ from scheme-A, 
compared with the experimental data from 
SLAC (large-$x$ data) \cite{Whitlow:1991uw},    
H1 and ZEUS collaborations (small-$x$ data) \cite{H1:2009pze}. 
One sees that the scheme-A predictions do not agree 
with the experimental data in both large-$x$ and small-$x$ regions, 
especially in the small-$x$ region. 
Scheme-A gives a way rapid growth of $F_2$ toward small $x$.  

In scheme-B, we implement the nonperturbative effects 
of DCSB and saturating $\alpha_{\rm s}$, 
but without the parton-parton recombination correction. 
In Figs. \ref{fig:F2_at_largex} and \ref{fig:F2_at_smallx}, 
we show the predicted $F_2$ at high $Q^2$ from scheme-B, 
compared with the experimental data from 
SLAC (large-$x$ data) \cite{Whitlow:1991uw},  H1 and ZEUS collaborations (small-$x$ data) \cite{H1:2009pze}.  
One also finds that the scheme-B predictions do not agree 
with the experimental data in both large-$x$ and small-$x$ regions, 
however they are much better than that from scheme-A. 
If we lower the hadronic scale for scheme-B, the consistency becomes better
in the large-$x$ region, however it becomes worse in small-$x$ region. 
If we increase the hadronic scale for scheme-B, the consistency becomes better 
in the small-$x$ region, however it becomes worse in large-$x$ region. 
We argue that considering the massive parton splitting, 
the parton evolution process slows down in the infrared region 
and the PDFs at small-$x$ is much lower. 
Nonetheless, the slope of PDF growing toward small-$x$ is still steeper 
than the measurements by H1 and ZEUS. 

Scheme-C is the proposed evolution scheme implemented 
with all corrections from DCSB, saturating strong coupling, 
and parton-parton recombination. 
In Figs. \ref{fig:F2_at_largex} and \ref{fig:F2_at_smallx}, 
we show the predicted $F_2$ at high $Q^2$ from scheme-C, 
compared with the experimental data from 
SLAC (large-$x$ data) \cite{Whitlow:1991uw},   H1 and ZEUS collaborations (small-$x$ data) \cite{H1:2009pze}.     
We find that the scheme-C predictions agree more or less 
with the experimental data in both large-$x$ and small-$x$ regions. 
Judged from the DIS data in the whole $x$ region down to $10^{-4}$, 
scheme-C surpass scheme-B and scheme-C. 
The scheme-C with the infrared modifications is an effective 
bridge for connecting the nucleon structures 
in the nonperturbative and perturbative domains. 

Comparing the calculations from the three different evolution schemes, 
we find that the nonperturbative and parton-parton recombination effects are crucial 
and indispensable for the DGLAP evolution at low $Q^2$ ($\lesssim M_{\rm q}^2$),  
in order to explain the nucleon structure data in the whole $x$ range. 
Firstly, the infrared-finite and saturating $\alpha_{\rm s}$ is a prominent 
nonperturbative phenomenon at low $Q^2$, 
which ensures the evolution equations are calculable 
from zero momentum scale. 
Secondly, DCSB dresses up the parton with a large effective mass, 
thus it slows down the parton splitting process with the factor $Q^2/(Q^2+M_q^2)$.  
As a result, the DGLAP evolution is frozen at extremely low $Q^2$ scale ($\ll M_q^2$).  
Although the parton effective mass suppresses the PDFs at small $x$ enormously, 
the PDF slope at small-$x$ is still quite large. 
Lastly, the parton-parton recombination correction reduces significantly 
the PDF slope at small $x$, giving a consistent result to the data. 
Due to the large parton size at low $Q^2$, 
the parton-parton recombination process can not be neglected. 

There are small discrepancies around $x\sim 0.01$ and in the large-$x$ region, 
in the comparisons between the scheme-C prediction and the data. 
The scheme-C prediction is lower than the data around $x\sim 0.01$, 
while it is higher than the data in the valence-quark region. 
The small discrepancies may be attributed to the fact that 
the MEM nonperturbative input is not that good. 
The addition of some intrinsic light quarks 
\cite{Brodsky:1980pb,Zou:2005xy,Zou:2010tc,Chang:2011vx,Liu:2012ch,Signal:1991ug,Melnitchouk:1998rv,Nikolaev:1998se} 
to the nonperturbative input 
may remove the small discrepancy around $x\sim 0.01$. 
At the same time, due to the momentum conservation, 
the valence quark distributions decrease a little bit, 
resulting a better consistency in the large-$x$ region as well.

\section{Summary}
\label{sec:summary}

We find an infrared-safe and effective evolution scheme 
for bridging the nucleon structures in nonperturbative and perturbative domains. 
The modified DGLAP evolution equations include the corrections 
from DCSB, saturating strong coupling and parton-parton recombination. 
Under the new evolution scheme, our results demonstrate that 
the naive quark model description of 
three valence quarks is an approximate origin 
of the PDFs accessed in the DIS measurement at the hard scale. 
The suggested evolution scheme reproduces the DIS data 
(SLAC, H1 and ZEUS) in the whole $x$ region  
from very small $x \sim 10^{-4}$. 

The emergent hadron mass mechanism and the nonlinear 
process of parton-parton recombination play crucial roles in 
DGLAP evolution at low momentum scales. 
The saturating $\alpha_{\rm s}$ due to an infrared gluon mass scale 
enables the calculability of the DGLAP equation from any low $Q^2$ scale. 
The parton effective mass from DCSB slows down the parton splitting 
in infrared region, leading to the freeze of PDF evolution at extremely low scales. 
This allows for a natural hadronic scale at the infrared limit, $Q_0^2=0$ GeV$^2$.  
The parton-parton recombination correction is a negative contribution 
to the growth of PDF at small $x$, 
reproducing the consistent PDF slope 
compared to the experimental measurements. 

At the end, we note that there remains room for refinement, 
particularly regarding the nonperturbative input at the hadronic scale 
and the evolution scheme in the infrared region. 
The small discrepancies between the theoretical prediction 
and the experimental data could be addressed 
by incorporating the intrinsic light quarks at the hadronic scale, 
potentially via the pion cloud model 
\cite{Signal:1991ug,Melnitchouk:1998rv,Nikolaev:1998se} 
or the five-quark component in the nucleon 
\cite{Zou:2010tc,Chang:2011vx,Liu:2012ch}. 
The suggested infrared-safe evolution scheme offers an effective tool 
for unifying the nucleon structures across the whole momentum scale. 
Nevertheless, this is just the beginning and preliminary test 
of extending the DGLAP evolution to far infrared region. 
More future studies are looking forward, 
regarding the nonperturbative and nonlinear effects 
to QCD evolution equations.

\begin{acknowledgments}
This work is supported by the National Natural Science Foundation of China
under the Grant NOs. 12005266 and 12305127,
the International Partnership Program of the Chinese Academy of Sciences under the Grant NO. 016GJHZ2022054FN,
and the Strategic Priority Research Program of Chinese Academy of Sciences under the Grant NO. XDB34030301.
\end{acknowledgments}

\bibliographystyle{apsrev4-1}
\bibliography{references}

\begin{thebibliography}{102}%
\makeatletter
\providecommand \@ifxundefined [1]{%
 \@ifx{#1\undefined}
}%
\providecommand \@ifnum [1]{%
 \ifnum #1\expandafter \@firstoftwo
 \else \expandafter \@secondoftwo
 \fi
}%
\providecommand \@ifx [1]{%
 \ifx #1\expandafter \@firstoftwo
 \else \expandafter \@secondoftwo
 \fi
}%
\providecommand \natexlab [1]{#1}%
\providecommand \enquote  [1]{``#1''}%
\providecommand \bibnamefont  [1]{#1}%
\providecommand \bibfnamefont [1]{#1}%
\providecommand \citenamefont [1]{#1}%
\providecommand \href@noop [0]{\@secondoftwo}%
\providecommand \href [0]{\begingroup \@sanitize@url \@href}%
\providecommand \@href[1]{\@@startlink{#1}\@@href}%
\providecommand \@@href[1]{\endgroup#1\@@endlink}%
\providecommand \@sanitize@url [0]{\catcode `\\12\catcode `\$12\catcode
  `\&12\catcode `\#12\catcode `\^12\catcode `\_12\catcode `\%12\relax}%
\providecommand \@@startlink[1]{}%
\providecommand \@@endlink[0]{}%
\providecommand \url  [0]{\begingroup\@sanitize@url \@url }%
\providecommand \@url [1]{\endgroup\@href {#1}{\urlprefix }}%
\providecommand \urlprefix  [0]{URL }%
\providecommand \Eprint [0]{\href }%
\providecommand \doibase [0]{http://dx.doi.org/}%
\providecommand \selectlanguage [0]{\@gobble}%
\providecommand \bibinfo  [0]{\@secondoftwo}%
\providecommand \bibfield  [0]{\@secondoftwo}%
\providecommand \translation [1]{[#1]}%
\providecommand \BibitemOpen [0]{}%
\providecommand \bibitemStop [0]{}%
\providecommand \bibitemNoStop [0]{.\EOS\space}%
\providecommand \EOS [0]{\spacefactor3000\relax}%
\providecommand \BibitemShut  [1]{\csname bibitem#1\endcsname}%
\let\auto@bib@innerbib\@empty
\bibitem [{\citenamefont {Wilczek}(2012)}]{Wilczek:2012sb}%
  \BibitemOpen
  \bibfield  {author} {\bibinfo {author} {\bibfnamefont {F.}~\bibnamefont
  {Wilczek}},\ }\href {\doibase 10.2478/s11534-012-0121-0} {\bibfield
  {journal} {\bibinfo  {journal} {Central Eur. J. Phys.}\ }\textbf {\bibinfo
  {volume} {10}},\ \bibinfo {pages} {1021} (\bibinfo {year} {2012})},\ \Eprint
  {http://arxiv.org/abs/1206.7114} {arXiv:1206.7114 [hep-ph]} \BibitemShut
  {NoStop}%
\bibitem [{\citenamefont {Ji}(1995)}]{Ji:1994av}%
  \BibitemOpen
  \bibfield  {author} {\bibinfo {author} {\bibfnamefont {X.-D.}\ \bibnamefont
  {Ji}},\ }\href {\doibase 10.1103/PhysRevLett.74.1071} {\bibfield  {journal}
  {\bibinfo  {journal} {Phys. Rev. Lett.}\ }\textbf {\bibinfo {volume} {74}},\
  \bibinfo {pages} {1071} (\bibinfo {year} {1995})},\ \Eprint
  {http://arxiv.org/abs/hep-ph/9410274} {arXiv:hep-ph/9410274} \BibitemShut
  {NoStop}%
\bibitem [{\citenamefont {Lorc\'e}(2018)}]{Lorce:2017xzd}%
  \BibitemOpen
  \bibfield  {author} {\bibinfo {author} {\bibfnamefont {C.}~\bibnamefont
  {Lorc\'e}},\ }\href {\doibase 10.1140/epjc/s10052-018-5561-2} {\bibfield
  {journal} {\bibinfo  {journal} {Eur. Phys. J. C}\ }\textbf {\bibinfo {volume}
  {78}},\ \bibinfo {pages} {120} (\bibinfo {year} {2018})},\ \Eprint
  {http://arxiv.org/abs/1706.05853} {arXiv:1706.05853 [hep-ph]} \BibitemShut
  {NoStop}%
\bibitem [{\citenamefont {Metz}\ \emph {et~al.}(2020)\citenamefont {Metz},
  \citenamefont {Pasquini},\ and\ \citenamefont {Rodini}}]{Metz:2020vxd}%
  \BibitemOpen
  \bibfield  {author} {\bibinfo {author} {\bibfnamefont {A.}~\bibnamefont
  {Metz}}, \bibinfo {author} {\bibfnamefont {B.}~\bibnamefont {Pasquini}}, \
  and\ \bibinfo {author} {\bibfnamefont {S.}~\bibnamefont {Rodini}},\ }\href
  {\doibase 10.1103/PhysRevD.102.114042} {\bibfield  {journal} {\bibinfo
  {journal} {Phys. Rev. D}\ }\textbf {\bibinfo {volume} {102}},\ \bibinfo
  {pages} {114042} (\bibinfo {year} {2020})},\ \Eprint
  {http://arxiv.org/abs/2006.11171} {arXiv:2006.11171 [hep-ph]} \BibitemShut
  {NoStop}%
\bibitem [{\citenamefont {Wang}\ \emph {et~al.}(2020)\citenamefont {Wang},
  \citenamefont {Evslin},\ and\ \citenamefont {Chen}}]{Wang:2019mza}%
  \BibitemOpen
  \bibfield  {author} {\bibinfo {author} {\bibfnamefont {R.}~\bibnamefont
  {Wang}}, \bibinfo {author} {\bibfnamefont {J.}~\bibnamefont {Evslin}}, \ and\
  \bibinfo {author} {\bibfnamefont {X.}~\bibnamefont {Chen}},\ }\href {\doibase
  10.1140/epjc/s10052-020-8057-9} {\bibfield  {journal} {\bibinfo  {journal}
  {Eur. Phys. J. C}\ }\textbf {\bibinfo {volume} {80}},\ \bibinfo {pages} {507}
  (\bibinfo {year} {2020})},\ \Eprint {http://arxiv.org/abs/1912.12040}
  {arXiv:1912.12040 [hep-ph]} \BibitemShut {NoStop}%
\bibitem [{\citenamefont {Jaffe}\ and\ \citenamefont
  {Manohar}(1990)}]{Jaffe:1989jz}%
  \BibitemOpen
  \bibfield  {author} {\bibinfo {author} {\bibfnamefont {R.~L.}\ \bibnamefont
  {Jaffe}}\ and\ \bibinfo {author} {\bibfnamefont {A.}~\bibnamefont
  {Manohar}},\ }\href {\doibase 10.1016/0550-3213(90)90506-9} {\bibfield
  {journal} {\bibinfo  {journal} {Nucl. Phys. B}\ }\textbf {\bibinfo {volume}
  {337}},\ \bibinfo {pages} {509} (\bibinfo {year} {1990})}\BibitemShut
  {NoStop}%
\bibitem [{\citenamefont {Ji}(1997)}]{Ji:1996ek}%
  \BibitemOpen
  \bibfield  {author} {\bibinfo {author} {\bibfnamefont {X.-D.}\ \bibnamefont
  {Ji}},\ }\href {\doibase 10.1103/PhysRevLett.78.610} {\bibfield  {journal}
  {\bibinfo  {journal} {Phys. Rev. Lett.}\ }\textbf {\bibinfo {volume} {78}},\
  \bibinfo {pages} {610} (\bibinfo {year} {1997})},\ \Eprint
  {http://arxiv.org/abs/hep-ph/9603249} {arXiv:hep-ph/9603249} \BibitemShut
  {NoStop}%
\bibitem [{\citenamefont {Ji}\ \emph {et~al.}(2021{\natexlab{a}})\citenamefont
  {Ji}, \citenamefont {Yuan},\ and\ \citenamefont {Zhao}}]{Ji:2020ena}%
  \BibitemOpen
  \bibfield  {author} {\bibinfo {author} {\bibfnamefont {X.}~\bibnamefont
  {Ji}}, \bibinfo {author} {\bibfnamefont {F.}~\bibnamefont {Yuan}}, \ and\
  \bibinfo {author} {\bibfnamefont {Y.}~\bibnamefont {Zhao}},\ }\href {\doibase
  10.1038/s42254-020-00248-4} {\bibfield  {journal} {\bibinfo  {journal}
  {Nature Rev. Phys.}\ }\textbf {\bibinfo {volume} {3}},\ \bibinfo {pages} {27}
  (\bibinfo {year} {2021}{\natexlab{a}})},\ \Eprint
  {http://arxiv.org/abs/2009.01291} {arXiv:2009.01291 [hep-ph]} \BibitemShut
  {NoStop}%
\bibitem [{\citenamefont {Gross}\ and\ \citenamefont
  {Wilczek}(1973)}]{Gross:1973id}%
  \BibitemOpen
  \bibfield  {author} {\bibinfo {author} {\bibfnamefont {D.~J.}\ \bibnamefont
  {Gross}}\ and\ \bibinfo {author} {\bibfnamefont {F.}~\bibnamefont
  {Wilczek}},\ }\href {\doibase 10.1103/PhysRevLett.30.1343} {\bibfield
  {journal} {\bibinfo  {journal} {Phys. Rev. Lett.}\ }\textbf {\bibinfo
  {volume} {30}},\ \bibinfo {pages} {1343} (\bibinfo {year}
  {1973})}\BibitemShut {NoStop}%
\bibitem [{\citenamefont {Politzer}(1973)}]{Politzer:1973fx}%
  \BibitemOpen
  \bibfield  {author} {\bibinfo {author} {\bibfnamefont {H.~D.}\ \bibnamefont
  {Politzer}},\ }\href {\doibase 10.1103/PhysRevLett.30.1346} {\bibfield
  {journal} {\bibinfo  {journal} {Phys. Rev. Lett.}\ }\textbf {\bibinfo
  {volume} {30}},\ \bibinfo {pages} {1346} (\bibinfo {year}
  {1973})}\BibitemShut {NoStop}%
\bibitem [{\citenamefont {Yang}\ and\ \citenamefont
  {Mills}(1954)}]{Yang:1954ek}%
  \BibitemOpen
  \bibfield  {author} {\bibinfo {author} {\bibfnamefont {C.-N.}\ \bibnamefont
  {Yang}}\ and\ \bibinfo {author} {\bibfnamefont {R.~L.}\ \bibnamefont
  {Mills}},\ }\href {\doibase 10.1103/PhysRev.96.191} {\bibfield  {journal}
  {\bibinfo  {journal} {Phys. Rev.}\ }\textbf {\bibinfo {volume} {96}},\
  \bibinfo {pages} {191} (\bibinfo {year} {1954})}\BibitemShut {NoStop}%
\bibitem [{\citenamefont {Fritzsch}\ \emph {et~al.}(1973)\citenamefont
  {Fritzsch}, \citenamefont {Gell-Mann},\ and\ \citenamefont
  {Leutwyler}}]{Fritzsch:1973pi}%
  \BibitemOpen
  \bibfield  {author} {\bibinfo {author} {\bibfnamefont {H.}~\bibnamefont
  {Fritzsch}}, \bibinfo {author} {\bibfnamefont {M.}~\bibnamefont {Gell-Mann}},
  \ and\ \bibinfo {author} {\bibfnamefont {H.}~\bibnamefont {Leutwyler}},\
  }\href {\doibase 10.1016/0370-2693(73)90625-4} {\bibfield  {journal}
  {\bibinfo  {journal} {Phys. Lett. B}\ }\textbf {\bibinfo {volume} {47}},\
  \bibinfo {pages} {365} (\bibinfo {year} {1973})}\BibitemShut {NoStop}%
\bibitem [{\citenamefont {'t~Hooft}\ and\ \citenamefont
  {Veltman}(1972)}]{tHooft:1972tcz}%
  \BibitemOpen
  \bibfield  {author} {\bibinfo {author} {\bibfnamefont {G.}~\bibnamefont
  {'t~Hooft}}\ and\ \bibinfo {author} {\bibfnamefont {M.~J.~G.}\ \bibnamefont
  {Veltman}},\ }\href {\doibase 10.1016/0550-3213(72)90279-9} {\bibfield
  {journal} {\bibinfo  {journal} {Nucl. Phys. B}\ }\textbf {\bibinfo {volume}
  {44}},\ \bibinfo {pages} {189} (\bibinfo {year} {1972})}\BibitemShut
  {NoStop}%
\bibitem [{\citenamefont {Ding}\ \emph {et~al.}(2023)\citenamefont {Ding},
  \citenamefont {Roberts},\ and\ \citenamefont {Schmidt}}]{Ding:2022ows}%
  \BibitemOpen
  \bibfield  {author} {\bibinfo {author} {\bibfnamefont {M.}~\bibnamefont
  {Ding}}, \bibinfo {author} {\bibfnamefont {C.~D.}\ \bibnamefont {Roberts}}, \
  and\ \bibinfo {author} {\bibfnamefont {S.~M.}\ \bibnamefont {Schmidt}},\
  }\href@noop {} {\bibfield  {journal} {\bibinfo  {journal} {Particles}\
  }\textbf {\bibinfo {volume} {6}},\ \bibinfo {pages} {57} (\bibinfo {year}
  {2023})},\ \Eprint {http://arxiv.org/abs/2211.07763} {arXiv:2211.07763
  [hep-ph]} \BibitemShut {NoStop}%
\bibitem [{\citenamefont {Papavassiliou}(2022)}]{Papavassiliou:2022wrb}%
  \BibitemOpen
  \bibfield  {author} {\bibinfo {author} {\bibfnamefont {J.}~\bibnamefont
  {Papavassiliou}},\ }\href {\doibase 10.1088/1674-1137/ac84ca} {\bibfield
  {journal} {\bibinfo  {journal} {Chin. Phys. C}\ }\textbf {\bibinfo {volume}
  {46}},\ \bibinfo {pages} {112001} (\bibinfo {year} {2022})},\ \Eprint
  {http://arxiv.org/abs/2207.04977} {arXiv:2207.04977 [hep-ph]} \BibitemShut
  {NoStop}%
\bibitem [{\citenamefont {Binosi}(2022)}]{Binosi:2022djx}%
  \BibitemOpen
  \bibfield  {author} {\bibinfo {author} {\bibfnamefont {D.}~\bibnamefont
  {Binosi}},\ }\href {\doibase 10.1007/s00601-022-01740-6} {\bibfield
  {journal} {\bibinfo  {journal} {Few Body Syst.}\ }\textbf {\bibinfo {volume}
  {63}},\ \bibinfo {pages} {42} (\bibinfo {year} {2022})},\ \Eprint
  {http://arxiv.org/abs/2203.00942} {arXiv:2203.00942 [hep-ph]} \BibitemShut
  {NoStop}%
\bibitem [{\citenamefont {Roberts}\ \emph {et~al.}(2021)\citenamefont
  {Roberts}, \citenamefont {Richards}, \citenamefont {Horn},\ and\
  \citenamefont {Chang}}]{Roberts:2021nhw}%
  \BibitemOpen
  \bibfield  {author} {\bibinfo {author} {\bibfnamefont {C.~D.}\ \bibnamefont
  {Roberts}}, \bibinfo {author} {\bibfnamefont {D.~G.}\ \bibnamefont
  {Richards}}, \bibinfo {author} {\bibfnamefont {T.}~\bibnamefont {Horn}}, \
  and\ \bibinfo {author} {\bibfnamefont {L.}~\bibnamefont {Chang}},\ }\href
  {\doibase 10.1016/j.ppnp.2021.103883} {\bibfield  {journal} {\bibinfo
  {journal} {Prog. Part. Nucl. Phys.}\ }\textbf {\bibinfo {volume} {120}},\
  \bibinfo {pages} {103883} (\bibinfo {year} {2021})},\ \Eprint
  {http://arxiv.org/abs/2102.01765} {arXiv:2102.01765 [hep-ph]} \BibitemShut
  {NoStop}%
\bibitem [{\citenamefont {Carman}\ \emph {et~al.}(2023)\citenamefont {Carman},
  \citenamefont {Gothe}, \citenamefont {Mokeev},\ and\ \citenamefont
  {Roberts}}]{Carman:2023zke}%
  \BibitemOpen
  \bibfield  {author} {\bibinfo {author} {\bibfnamefont {D.~S.}\ \bibnamefont
  {Carman}}, \bibinfo {author} {\bibfnamefont {R.~W.}\ \bibnamefont {Gothe}},
  \bibinfo {author} {\bibfnamefont {V.~I.}\ \bibnamefont {Mokeev}}, \ and\
  \bibinfo {author} {\bibfnamefont {C.~D.}\ \bibnamefont {Roberts}},\ }\href
  {\doibase 10.3390/particles6010023} {\bibfield  {journal} {\bibinfo
  {journal} {Particles}\ }\textbf {\bibinfo {volume} {6}},\ \bibinfo {pages}
  {416} (\bibinfo {year} {2023})},\ \Eprint {http://arxiv.org/abs/2301.07777}
  {arXiv:2301.07777 [hep-ph]} \BibitemShut {NoStop}%
\bibitem [{\citenamefont {Abdul~Khalek}\ \emph {et~al.}(2022)\citenamefont
  {Abdul~Khalek} \emph {et~al.}}]{AbdulKhalek:2021gbh}%
  \BibitemOpen
  \bibfield  {author} {\bibinfo {author} {\bibfnamefont {R.}~\bibnamefont
  {Abdul~Khalek}} \emph {et~al.},\ }\href {\doibase
  10.1016/j.nuclphysa.2022.122447} {\bibfield  {journal} {\bibinfo  {journal}
  {Nucl. Phys. A}\ }\textbf {\bibinfo {volume} {1026}},\ \bibinfo {pages}
  {122447} (\bibinfo {year} {2022})},\ \Eprint
  {http://arxiv.org/abs/2103.05419} {arXiv:2103.05419 [physics.ins-det]}
  \BibitemShut {NoStop}%
\bibitem [{\citenamefont {Accardi}\ \emph {et~al.}(2016)\citenamefont {Accardi}
  \emph {et~al.}}]{Accardi:2012qut}%
  \BibitemOpen
  \bibfield  {author} {\bibinfo {author} {\bibfnamefont {A.}~\bibnamefont
  {Accardi}} \emph {et~al.},\ }\href {\doibase 10.1140/epja/i2016-16268-9}
  {\bibfield  {journal} {\bibinfo  {journal} {Eur. Phys. J. A}\ }\textbf
  {\bibinfo {volume} {52}},\ \bibinfo {pages} {268} (\bibinfo {year} {2016})},\
  \Eprint {http://arxiv.org/abs/1212.1701} {arXiv:1212.1701 [nucl-ex]}
  \BibitemShut {NoStop}%
\bibitem [{\citenamefont {Anderle}\ \emph {et~al.}(2021)\citenamefont {Anderle}
  \emph {et~al.}}]{Anderle:2021wcy}%
  \BibitemOpen
  \bibfield  {author} {\bibinfo {author} {\bibfnamefont {D.~P.}\ \bibnamefont
  {Anderle}} \emph {et~al.},\ }\href {\doibase 10.1007/s11467-021-1062-0}
  {\bibfield  {journal} {\bibinfo  {journal} {Front. Phys. (Beijing)}\ }\textbf
  {\bibinfo {volume} {16}},\ \bibinfo {pages} {64701} (\bibinfo {year}
  {2021})},\ \Eprint {http://arxiv.org/abs/2102.09222} {arXiv:2102.09222
  [nucl-ex]} \BibitemShut {NoStop}%
\bibitem [{\citenamefont {Chen}(2018)}]{Chen:2018wyz}%
  \BibitemOpen
  \bibfield  {author} {\bibinfo {author} {\bibfnamefont {X.}~\bibnamefont
  {Chen}},\ }\href {\doibase 10.22323/1.316.0170} {\bibfield  {journal}
  {\bibinfo  {journal} {PoS}\ }\textbf {\bibinfo {volume} {DIS2018}},\ \bibinfo
  {pages} {170} (\bibinfo {year} {2018})},\ \Eprint
  {http://arxiv.org/abs/1809.00448} {arXiv:1809.00448 [nucl-ex]} \BibitemShut
  {NoStop}%
\bibitem [{\citenamefont {Chen}\ \emph {et~al.}(2020)\citenamefont {Chen},
  \citenamefont {Guo}, \citenamefont {Roberts},\ and\ \citenamefont
  {Wang}}]{Chen:2020ijn}%
  \BibitemOpen
  \bibfield  {author} {\bibinfo {author} {\bibfnamefont {X.}~\bibnamefont
  {Chen}}, \bibinfo {author} {\bibfnamefont {F.-K.}\ \bibnamefont {Guo}},
  \bibinfo {author} {\bibfnamefont {C.~D.}\ \bibnamefont {Roberts}}, \ and\
  \bibinfo {author} {\bibfnamefont {R.}~\bibnamefont {Wang}},\ }\href {\doibase
  10.1007/s00601-020-01574-0} {\bibfield  {journal} {\bibinfo  {journal} {Few
  Body Syst.}\ }\textbf {\bibinfo {volume} {61}},\ \bibinfo {pages} {43}
  (\bibinfo {year} {2020})},\ \Eprint {http://arxiv.org/abs/2008.00102}
  {arXiv:2008.00102 [hep-ph]} \BibitemShut {NoStop}%
\bibitem [{\citenamefont {Wang}\ and\ \citenamefont
  {Chen}(2022)}]{Wang:2022xad}%
  \BibitemOpen
  \bibfield  {author} {\bibinfo {author} {\bibfnamefont {R.}~\bibnamefont
  {Wang}}\ and\ \bibinfo {author} {\bibfnamefont {X.}~\bibnamefont {Chen}},\
  }\href {\doibase 10.1007/s00601-022-01751-3} {\bibfield  {journal} {\bibinfo
  {journal} {Few Body Syst.}\ }\textbf {\bibinfo {volume} {63}},\ \bibinfo
  {pages} {48} (\bibinfo {year} {2022})}\BibitemShut {NoStop}%
\bibitem [{\citenamefont {Accardi}\ \emph {et~al.}(2023)\citenamefont {Accardi}
  \emph {et~al.}}]{Accardi:2023chb}%
  \BibitemOpen
  \bibfield  {author} {\bibinfo {author} {\bibfnamefont {A.}~\bibnamefont
  {Accardi}} \emph {et~al.},\ }\href@noop {} {\  (\bibinfo {year} {2023})},\
  \Eprint {http://arxiv.org/abs/2306.09360} {arXiv:2306.09360 [nucl-ex]}
  \BibitemShut {NoStop}%
\bibitem [{\citenamefont {Workman}\ \emph {et~al.}(2022)\citenamefont {Workman}
  \emph {et~al.}}]{ParticleDataGroup:2022pth}%
  \BibitemOpen
  \bibfield  {author} {\bibinfo {author} {\bibfnamefont {R.~L.}\ \bibnamefont
  {Workman}} \emph {et~al.} (\bibinfo {collaboration} {Particle Data Group}),\
  }\href {\doibase 10.1093/ptep/ptac097} {\bibfield  {journal} {\bibinfo
  {journal} {PTEP}\ }\textbf {\bibinfo {volume} {2022}},\ \bibinfo {pages}
  {083C01} (\bibinfo {year} {2022})}\BibitemShut {NoStop}%
\bibitem [{\citenamefont {Bjorken}(1969)}]{Bjorken:1968dy}%
  \BibitemOpen
  \bibfield  {author} {\bibinfo {author} {\bibfnamefont {J.~D.}\ \bibnamefont
  {Bjorken}},\ }\href {\doibase 10.1103/PhysRev.179.1547} {\bibfield  {journal}
  {\bibinfo  {journal} {Phys. Rev.}\ }\textbf {\bibinfo {volume} {179}},\
  \bibinfo {pages} {1547} (\bibinfo {year} {1969})}\BibitemShut {NoStop}%
\bibitem [{\citenamefont {Feynman}(1969)}]{Feynman:1969ej}%
  \BibitemOpen
  \bibfield  {author} {\bibinfo {author} {\bibfnamefont {R.~P.}\ \bibnamefont
  {Feynman}},\ }\href {\doibase 10.1103/PhysRevLett.23.1415} {\bibfield
  {journal} {\bibinfo  {journal} {Phys. Rev. Lett.}\ }\textbf {\bibinfo
  {volume} {23}},\ \bibinfo {pages} {1415} (\bibinfo {year}
  {1969})}\BibitemShut {NoStop}%
\bibitem [{\citenamefont {Bjorken}\ and\ \citenamefont
  {Paschos}(1969)}]{Bjorken:1969ja}%
  \BibitemOpen
  \bibfield  {author} {\bibinfo {author} {\bibfnamefont {J.~D.}\ \bibnamefont
  {Bjorken}}\ and\ \bibinfo {author} {\bibfnamefont {E.~A.}\ \bibnamefont
  {Paschos}},\ }\href {\doibase 10.1103/PhysRev.185.1975} {\bibfield  {journal}
  {\bibinfo  {journal} {Phys. Rev.}\ }\textbf {\bibinfo {volume} {185}},\
  \bibinfo {pages} {1975} (\bibinfo {year} {1969})}\BibitemShut {NoStop}%
\bibitem [{\citenamefont {Collins}\ and\ \citenamefont
  {Soper}(1987)}]{Collins:1987pm}%
  \BibitemOpen
  \bibfield  {author} {\bibinfo {author} {\bibfnamefont {J.~C.}\ \bibnamefont
  {Collins}}\ and\ \bibinfo {author} {\bibfnamefont {D.~E.}\ \bibnamefont
  {Soper}},\ }\href {\doibase 10.1146/annurev.ns.37.120187.002123} {\bibfield
  {journal} {\bibinfo  {journal} {Ann. Rev. Nucl. Part. Sci.}\ }\textbf
  {\bibinfo {volume} {37}},\ \bibinfo {pages} {383} (\bibinfo {year}
  {1987})}\BibitemShut {NoStop}%
\bibitem [{\citenamefont {Collins}\ \emph {et~al.}(1989)\citenamefont
  {Collins}, \citenamefont {Soper},\ and\ \citenamefont
  {Sterman}}]{Collins:1989gx}%
  \BibitemOpen
  \bibfield  {author} {\bibinfo {author} {\bibfnamefont {J.~C.}\ \bibnamefont
  {Collins}}, \bibinfo {author} {\bibfnamefont {D.~E.}\ \bibnamefont {Soper}},
  \ and\ \bibinfo {author} {\bibfnamefont {G.~F.}\ \bibnamefont {Sterman}},\
  }\href {\doibase 10.1142/9789814503266_0001} {\bibfield  {journal} {\bibinfo
  {journal} {Adv. Ser. Direct. High Energy Phys.}\ }\textbf {\bibinfo {volume}
  {5}},\ \bibinfo {pages} {1} (\bibinfo {year} {1989})},\ \Eprint
  {http://arxiv.org/abs/hep-ph/0409313} {arXiv:hep-ph/0409313} \BibitemShut
  {NoStop}%
\bibitem [{\citenamefont {Sterman}(1995)}]{Sterman:1995fz}%
  \BibitemOpen
  \bibfield  {author} {\bibinfo {author} {\bibfnamefont {G.~F.}\ \bibnamefont
  {Sterman}},\ }in\ \href@noop {} {\emph {\bibinfo {booktitle} {{Theoretical
  Advanced Study Institute in Elementary Particle Physics (TASI 95): QCD and
  Beyond}}}}\ (\bibinfo {year} {1995})\ pp.\ \bibinfo {pages} {327--408},\
  \Eprint {http://arxiv.org/abs/hep-ph/9606312} {arXiv:hep-ph/9606312}
  \BibitemShut {NoStop}%
\bibitem [{\citenamefont {Taylor}(1991)}]{Taylor:1991ew}%
  \BibitemOpen
  \bibfield  {author} {\bibinfo {author} {\bibfnamefont {R.~E.}\ \bibnamefont
  {Taylor}},\ }\href {\doibase 10.1103/RevModPhys.63.573} {\bibfield  {journal}
  {\bibinfo  {journal} {Rev. Mod. Phys.}\ }\textbf {\bibinfo {volume} {63}},\
  \bibinfo {pages} {573} (\bibinfo {year} {1991})}\BibitemShut {NoStop}%
\bibitem [{\citenamefont {Kendall}(1991)}]{Kendall:1991np}%
  \BibitemOpen
  \bibfield  {author} {\bibinfo {author} {\bibfnamefont {H.~W.}\ \bibnamefont
  {Kendall}},\ }\href {\doibase 10.1103/RevModPhys.63.597} {\bibfield
  {journal} {\bibinfo  {journal} {Rev. Mod. Phys.}\ }\textbf {\bibinfo {volume}
  {63}},\ \bibinfo {pages} {597} (\bibinfo {year} {1991})}\BibitemShut
  {NoStop}%
\bibitem [{\citenamefont {Friedman}(1991)}]{Friedman:1991nq}%
  \BibitemOpen
  \bibfield  {author} {\bibinfo {author} {\bibfnamefont {J.~I.}\ \bibnamefont
  {Friedman}},\ }\href {\doibase 10.1103/RevModPhys.63.615} {\bibfield
  {journal} {\bibinfo  {journal} {Rev. Mod. Phys.}\ }\textbf {\bibinfo {volume}
  {63}},\ \bibinfo {pages} {615} (\bibinfo {year} {1991})}\BibitemShut
  {NoStop}%
\bibitem [{\citenamefont {Pumplin}\ \emph {et~al.}(2002)\citenamefont
  {Pumplin}, \citenamefont {Stump}, \citenamefont {Huston}, \citenamefont
  {Lai}, \citenamefont {Nadolsky},\ and\ \citenamefont
  {Tung}}]{Pumplin:2002vw}%
  \BibitemOpen
  \bibfield  {author} {\bibinfo {author} {\bibfnamefont {J.}~\bibnamefont
  {Pumplin}}, \bibinfo {author} {\bibfnamefont {D.~R.}\ \bibnamefont {Stump}},
  \bibinfo {author} {\bibfnamefont {J.}~\bibnamefont {Huston}}, \bibinfo
  {author} {\bibfnamefont {H.~L.}\ \bibnamefont {Lai}}, \bibinfo {author}
  {\bibfnamefont {P.~M.}\ \bibnamefont {Nadolsky}}, \ and\ \bibinfo {author}
  {\bibfnamefont {W.~K.}\ \bibnamefont {Tung}},\ }\href {\doibase
  10.1088/1126-6708/2002/07/012} {\bibfield  {journal} {\bibinfo  {journal}
  {JHEP}\ }\textbf {\bibinfo {volume} {07}},\ \bibinfo {pages} {012} (\bibinfo
  {year} {2002})},\ \Eprint {http://arxiv.org/abs/hep-ph/0201195}
  {arXiv:hep-ph/0201195} \BibitemShut {NoStop}%
\bibitem [{\citenamefont {Dulat}\ \emph {et~al.}(2016)\citenamefont {Dulat},
  \citenamefont {Hou}, \citenamefont {Gao}, \citenamefont {Guzzi},
  \citenamefont {Huston}, \citenamefont {Nadolsky}, \citenamefont {Pumplin},
  \citenamefont {Schmidt}, \citenamefont {Stump},\ and\ \citenamefont
  {Yuan}}]{Dulat:2015mca}%
  \BibitemOpen
  \bibfield  {author} {\bibinfo {author} {\bibfnamefont {S.}~\bibnamefont
  {Dulat}}, \bibinfo {author} {\bibfnamefont {T.-J.}\ \bibnamefont {Hou}},
  \bibinfo {author} {\bibfnamefont {J.}~\bibnamefont {Gao}}, \bibinfo {author}
  {\bibfnamefont {M.}~\bibnamefont {Guzzi}}, \bibinfo {author} {\bibfnamefont
  {J.}~\bibnamefont {Huston}}, \bibinfo {author} {\bibfnamefont
  {P.}~\bibnamefont {Nadolsky}}, \bibinfo {author} {\bibfnamefont
  {J.}~\bibnamefont {Pumplin}}, \bibinfo {author} {\bibfnamefont
  {C.}~\bibnamefont {Schmidt}}, \bibinfo {author} {\bibfnamefont
  {D.}~\bibnamefont {Stump}}, \ and\ \bibinfo {author} {\bibfnamefont {C.~P.}\
  \bibnamefont {Yuan}},\ }\href {\doibase 10.1103/PhysRevD.93.033006}
  {\bibfield  {journal} {\bibinfo  {journal} {Phys. Rev. D}\ }\textbf {\bibinfo
  {volume} {93}},\ \bibinfo {pages} {033006} (\bibinfo {year} {2016})},\
  \Eprint {http://arxiv.org/abs/1506.07443} {arXiv:1506.07443 [hep-ph]}
  \BibitemShut {NoStop}%
\bibitem [{\citenamefont {Ball}\ \emph {et~al.}(2012)\citenamefont {Ball},
  \citenamefont {Bertone}, \citenamefont {Cerutti}, \citenamefont {Del~Debbio},
  \citenamefont {Forte}, \citenamefont {Guffanti}, \citenamefont {Latorre},
  \citenamefont {Rojo},\ and\ \citenamefont {Ubiali}}]{Ball:2011uy}%
  \BibitemOpen
  \bibfield  {author} {\bibinfo {author} {\bibfnamefont {R.~D.}\ \bibnamefont
  {Ball}}, \bibinfo {author} {\bibfnamefont {V.}~\bibnamefont {Bertone}},
  \bibinfo {author} {\bibfnamefont {F.}~\bibnamefont {Cerutti}}, \bibinfo
  {author} {\bibfnamefont {L.}~\bibnamefont {Del~Debbio}}, \bibinfo {author}
  {\bibfnamefont {S.}~\bibnamefont {Forte}}, \bibinfo {author} {\bibfnamefont
  {A.}~\bibnamefont {Guffanti}}, \bibinfo {author} {\bibfnamefont {J.~I.}\
  \bibnamefont {Latorre}}, \bibinfo {author} {\bibfnamefont {J.}~\bibnamefont
  {Rojo}}, \ and\ \bibinfo {author} {\bibfnamefont {M.}~\bibnamefont {Ubiali}}
  (\bibinfo {collaboration} {NNPDF}),\ }\href {\doibase
  10.1016/j.nuclphysb.2011.09.024} {\bibfield  {journal} {\bibinfo  {journal}
  {Nucl. Phys. B}\ }\textbf {\bibinfo {volume} {855}},\ \bibinfo {pages} {153}
  (\bibinfo {year} {2012})},\ \Eprint {http://arxiv.org/abs/1107.2652}
  {arXiv:1107.2652 [hep-ph]} \BibitemShut {NoStop}%
\bibitem [{\citenamefont {Yu}\ \emph {et~al.}(2024)\citenamefont {Yu},
  \citenamefont {Cheng}, \citenamefont {Xing}, \citenamefont {Gao},\ and\
  \citenamefont {Roberts}}]{Yu:2024qsd}%
  \BibitemOpen
  \bibfield  {author} {\bibinfo {author} {\bibfnamefont {Y.}~\bibnamefont
  {Yu}}, \bibinfo {author} {\bibfnamefont {P.}~\bibnamefont {Cheng}}, \bibinfo
  {author} {\bibfnamefont {H.-Y.}\ \bibnamefont {Xing}}, \bibinfo {author}
  {\bibfnamefont {F.}~\bibnamefont {Gao}}, \ and\ \bibinfo {author}
  {\bibfnamefont {C.~D.}\ \bibnamefont {Roberts}},\ }\href@noop {} {\
  (\bibinfo {year} {2024})},\ \Eprint {http://arxiv.org/abs/2402.06095}
  {arXiv:2402.06095 [hep-ph]} \BibitemShut {NoStop}%
\bibitem [{\citenamefont {Lu}\ \emph {et~al.}(2022)\citenamefont {Lu},
  \citenamefont {Chang}, \citenamefont {Raya}, \citenamefont {Roberts},\ and\
  \citenamefont {Rodr\'\i{}guez-Quintero}}]{Lu:2022cjx}%
  \BibitemOpen
  \bibfield  {author} {\bibinfo {author} {\bibfnamefont {Y.}~\bibnamefont
  {Lu}}, \bibinfo {author} {\bibfnamefont {L.}~\bibnamefont {Chang}}, \bibinfo
  {author} {\bibfnamefont {K.}~\bibnamefont {Raya}}, \bibinfo {author}
  {\bibfnamefont {C.~D.}\ \bibnamefont {Roberts}}, \ and\ \bibinfo {author}
  {\bibfnamefont {J.}~\bibnamefont {Rodr\'\i{}guez-Quintero}},\ }\href
  {\doibase 10.1016/j.physletb.2022.137130} {\bibfield  {journal} {\bibinfo
  {journal} {Phys. Lett. B}\ }\textbf {\bibinfo {volume} {830}},\ \bibinfo
  {pages} {137130} (\bibinfo {year} {2022})},\ \Eprint
  {http://arxiv.org/abs/2203.00753} {arXiv:2203.00753 [hep-ph]} \BibitemShut
  {NoStop}%
\bibitem [{\citenamefont {Roberts}(2023)}]{Roberts:2023lap}%
  \BibitemOpen
  \bibfield  {author} {\bibinfo {author} {\bibfnamefont {C.~D.}\ \bibnamefont
  {Roberts}},\ }\href {\doibase 10.1007/s00601-023-01837-6} {\bibfield
  {journal} {\bibinfo  {journal} {Few Body Syst.}\ }\textbf {\bibinfo {volume}
  {64}},\ \bibinfo {pages} {51} (\bibinfo {year} {2023})},\ \Eprint
  {http://arxiv.org/abs/2304.00154} {arXiv:2304.00154 [hep-ph]} \BibitemShut
  {NoStop}%
\bibitem [{\citenamefont {Chang}\ \emph {et~al.}(2014)\citenamefont {Chang},
  \citenamefont {Mezrag}, \citenamefont {Moutarde}, \citenamefont {Roberts},
  \citenamefont {Rodr\'\i{}guez-Quintero},\ and\ \citenamefont
  {Tandy}}]{Chang:2014lva}%
  \BibitemOpen
  \bibfield  {author} {\bibinfo {author} {\bibfnamefont {L.}~\bibnamefont
  {Chang}}, \bibinfo {author} {\bibfnamefont {C.}~\bibnamefont {Mezrag}},
  \bibinfo {author} {\bibfnamefont {H.}~\bibnamefont {Moutarde}}, \bibinfo
  {author} {\bibfnamefont {C.~D.}\ \bibnamefont {Roberts}}, \bibinfo {author}
  {\bibfnamefont {J.}~\bibnamefont {Rodr\'\i{}guez-Quintero}}, \ and\ \bibinfo
  {author} {\bibfnamefont {P.~C.}\ \bibnamefont {Tandy}},\ }\href {\doibase
  10.1016/j.physletb.2014.08.009} {\bibfield  {journal} {\bibinfo  {journal}
  {Phys. Lett. B}\ }\textbf {\bibinfo {volume} {737}},\ \bibinfo {pages} {23}
  (\bibinfo {year} {2014})},\ \Eprint {http://arxiv.org/abs/1406.5450}
  {arXiv:1406.5450 [nucl-th]} \BibitemShut {NoStop}%
\bibitem [{\citenamefont {Chang}\ and\ \citenamefont
  {Thomas}(2015)}]{Chang:2014gga}%
  \BibitemOpen
  \bibfield  {author} {\bibinfo {author} {\bibfnamefont {L.}~\bibnamefont
  {Chang}}\ and\ \bibinfo {author} {\bibfnamefont {A.~W.}\ \bibnamefont
  {Thomas}},\ }\href {\doibase 10.1016/j.physletb.2015.08.036} {\bibfield
  {journal} {\bibinfo  {journal} {Phys. Lett. B}\ }\textbf {\bibinfo {volume}
  {749}},\ \bibinfo {pages} {547} (\bibinfo {year} {2015})},\ \Eprint
  {http://arxiv.org/abs/1410.8250} {arXiv:1410.8250 [nucl-th]} \BibitemShut
  {NoStop}%
\bibitem [{\citenamefont {Ding}\ \emph {et~al.}(2020)\citenamefont {Ding},
  \citenamefont {Raya}, \citenamefont {Binosi}, \citenamefont {Chang},
  \citenamefont {Roberts},\ and\ \citenamefont {Schmidt}}]{Ding:2019lwe}%
  \BibitemOpen
  \bibfield  {author} {\bibinfo {author} {\bibfnamefont {M.}~\bibnamefont
  {Ding}}, \bibinfo {author} {\bibfnamefont {K.}~\bibnamefont {Raya}}, \bibinfo
  {author} {\bibfnamefont {D.}~\bibnamefont {Binosi}}, \bibinfo {author}
  {\bibfnamefont {L.}~\bibnamefont {Chang}}, \bibinfo {author} {\bibfnamefont
  {C.~D.}\ \bibnamefont {Roberts}}, \ and\ \bibinfo {author} {\bibfnamefont
  {S.~M.}\ \bibnamefont {Schmidt}},\ }\href {\doibase
  10.1103/PhysRevD.101.054014} {\bibfield  {journal} {\bibinfo  {journal}
  {Phys. Rev. D}\ }\textbf {\bibinfo {volume} {101}},\ \bibinfo {pages}
  {054014} (\bibinfo {year} {2020})},\ \Eprint
  {http://arxiv.org/abs/1905.05208} {arXiv:1905.05208 [nucl-th]} \BibitemShut
  {NoStop}%
\bibitem [{\citenamefont {Cui}\ \emph {et~al.}(2020{\natexlab{a}})\citenamefont
  {Cui}, \citenamefont {Ding}, \citenamefont {Gao}, \citenamefont {Raya},
  \citenamefont {Binosi}, \citenamefont {Chang}, \citenamefont {Roberts},
  \citenamefont {Rodr\'\i{}guez-Quintero},\ and\ \citenamefont
  {Schmidt}}]{Cui:2020tdf}%
  \BibitemOpen
  \bibfield  {author} {\bibinfo {author} {\bibfnamefont {Z.-F.}\ \bibnamefont
  {Cui}}, \bibinfo {author} {\bibfnamefont {M.}~\bibnamefont {Ding}}, \bibinfo
  {author} {\bibfnamefont {F.}~\bibnamefont {Gao}}, \bibinfo {author}
  {\bibfnamefont {K.}~\bibnamefont {Raya}}, \bibinfo {author} {\bibfnamefont
  {D.}~\bibnamefont {Binosi}}, \bibinfo {author} {\bibfnamefont
  {L.}~\bibnamefont {Chang}}, \bibinfo {author} {\bibfnamefont {C.~D.}\
  \bibnamefont {Roberts}}, \bibinfo {author} {\bibfnamefont {J.}~\bibnamefont
  {Rodr\'\i{}guez-Quintero}}, \ and\ \bibinfo {author} {\bibfnamefont {S.~M.}\
  \bibnamefont {Schmidt}},\ }\href {\doibase 10.1140/epjc/s10052-020-08578-4}
  {\bibfield  {journal} {\bibinfo  {journal} {Eur. Phys. J. C}\ }\textbf
  {\bibinfo {volume} {80}},\ \bibinfo {pages} {1064} (\bibinfo {year}
  {2020}{\natexlab{a}})}\BibitemShut {NoStop}%
\bibitem [{\citenamefont {Ji}(2013)}]{Ji:2013dva}%
  \BibitemOpen
  \bibfield  {author} {\bibinfo {author} {\bibfnamefont {X.}~\bibnamefont
  {Ji}},\ }\href {\doibase 10.1103/PhysRevLett.110.262002} {\bibfield
  {journal} {\bibinfo  {journal} {Phys. Rev. Lett.}\ }\textbf {\bibinfo
  {volume} {110}},\ \bibinfo {pages} {262002} (\bibinfo {year} {2013})},\
  \Eprint {http://arxiv.org/abs/1305.1539} {arXiv:1305.1539 [hep-ph]}
  \BibitemShut {NoStop}%
\bibitem [{\citenamefont {Ji}(2014)}]{Ji:2014gla}%
  \BibitemOpen
  \bibfield  {author} {\bibinfo {author} {\bibfnamefont {X.}~\bibnamefont
  {Ji}},\ }\href {\doibase 10.1007/s11433-014-5492-3} {\bibfield  {journal}
  {\bibinfo  {journal} {Sci. China Phys. Mech. Astron.}\ }\textbf {\bibinfo
  {volume} {57}},\ \bibinfo {pages} {1407} (\bibinfo {year} {2014})},\ \Eprint
  {http://arxiv.org/abs/1404.6680} {arXiv:1404.6680 [hep-ph]} \BibitemShut
  {NoStop}%
\bibitem [{\citenamefont {Ji}\ \emph {et~al.}(2021{\natexlab{b}})\citenamefont
  {Ji}, \citenamefont {Liu}, \citenamefont {Liu}, \citenamefont {Zhang},\ and\
  \citenamefont {Zhao}}]{Ji:2020ect}%
  \BibitemOpen
  \bibfield  {author} {\bibinfo {author} {\bibfnamefont {X.}~\bibnamefont
  {Ji}}, \bibinfo {author} {\bibfnamefont {Y.-S.}\ \bibnamefont {Liu}},
  \bibinfo {author} {\bibfnamefont {Y.}~\bibnamefont {Liu}}, \bibinfo {author}
  {\bibfnamefont {J.-H.}\ \bibnamefont {Zhang}}, \ and\ \bibinfo {author}
  {\bibfnamefont {Y.}~\bibnamefont {Zhao}},\ }\href {\doibase
  10.1103/RevModPhys.93.035005} {\bibfield  {journal} {\bibinfo  {journal}
  {Rev. Mod. Phys.}\ }\textbf {\bibinfo {volume} {93}},\ \bibinfo {pages}
  {035005} (\bibinfo {year} {2021}{\natexlab{b}})},\ \Eprint
  {http://arxiv.org/abs/2004.03543} {arXiv:2004.03543 [hep-ph]} \BibitemShut
  {NoStop}%
\bibitem [{\citenamefont {Ji}\ \emph {et~al.}(2018)\citenamefont {Ji},
  \citenamefont {Zhang},\ and\ \citenamefont {Zhao}}]{Ji:2017oey}%
  \BibitemOpen
  \bibfield  {author} {\bibinfo {author} {\bibfnamefont {X.}~\bibnamefont
  {Ji}}, \bibinfo {author} {\bibfnamefont {J.-H.}\ \bibnamefont {Zhang}}, \
  and\ \bibinfo {author} {\bibfnamefont {Y.}~\bibnamefont {Zhao}},\ }\href
  {\doibase 10.1103/PhysRevLett.120.112001} {\bibfield  {journal} {\bibinfo
  {journal} {Phys. Rev. Lett.}\ }\textbf {\bibinfo {volume} {120}},\ \bibinfo
  {pages} {112001} (\bibinfo {year} {2018})},\ \Eprint
  {http://arxiv.org/abs/1706.08962} {arXiv:1706.08962 [hep-ph]} \BibitemShut
  {NoStop}%
\bibitem [{\citenamefont {Ma}\ and\ \citenamefont
  {Qiu}(2018{\natexlab{a}})}]{Ma:2014jla}%
  \BibitemOpen
  \bibfield  {author} {\bibinfo {author} {\bibfnamefont {Y.-Q.}\ \bibnamefont
  {Ma}}\ and\ \bibinfo {author} {\bibfnamefont {J.-W.}\ \bibnamefont {Qiu}},\
  }\href {\doibase 10.1103/PhysRevD.98.074021} {\bibfield  {journal} {\bibinfo
  {journal} {Phys. Rev. D}\ }\textbf {\bibinfo {volume} {98}},\ \bibinfo
  {pages} {074021} (\bibinfo {year} {2018}{\natexlab{a}})},\ \Eprint
  {http://arxiv.org/abs/1404.6860} {arXiv:1404.6860 [hep-ph]} \BibitemShut
  {NoStop}%
\bibitem [{\citenamefont {Ma}\ and\ \citenamefont
  {Qiu}(2018{\natexlab{b}})}]{Ma:2017pxb}%
  \BibitemOpen
  \bibfield  {author} {\bibinfo {author} {\bibfnamefont {Y.-Q.}\ \bibnamefont
  {Ma}}\ and\ \bibinfo {author} {\bibfnamefont {J.-W.}\ \bibnamefont {Qiu}},\
  }\href {\doibase 10.1103/PhysRevLett.120.022003} {\bibfield  {journal}
  {\bibinfo  {journal} {Phys. Rev. Lett.}\ }\textbf {\bibinfo {volume} {120}},\
  \bibinfo {pages} {022003} (\bibinfo {year} {2018}{\natexlab{b}})},\ \Eprint
  {http://arxiv.org/abs/1709.03018} {arXiv:1709.03018 [hep-ph]} \BibitemShut
  {NoStop}%
\bibitem [{\citenamefont {Constantinou}\ \emph {et~al.}(2021)\citenamefont
  {Constantinou} \emph {et~al.}}]{Constantinou:2020hdm}%
  \BibitemOpen
  \bibfield  {author} {\bibinfo {author} {\bibfnamefont {M.}~\bibnamefont
  {Constantinou}} \emph {et~al.},\ }\href {\doibase 10.1016/j.ppnp.2021.103908}
  {\bibfield  {journal} {\bibinfo  {journal} {Prog. Part. Nucl. Phys.}\
  }\textbf {\bibinfo {volume} {121}},\ \bibinfo {pages} {103908} (\bibinfo
  {year} {2021})},\ \Eprint {http://arxiv.org/abs/2006.08636} {arXiv:2006.08636
  [hep-ph]} \BibitemShut {NoStop}%
\bibitem [{\citenamefont {Lin}\ \emph {et~al.}(2018)\citenamefont {Lin} \emph
  {et~al.}}]{Lin:2017snn}%
  \BibitemOpen
  \bibfield  {author} {\bibinfo {author} {\bibfnamefont {H.-W.}\ \bibnamefont
  {Lin}} \emph {et~al.},\ }\href {\doibase 10.1016/j.ppnp.2018.01.007}
  {\bibfield  {journal} {\bibinfo  {journal} {Prog. Part. Nucl. Phys.}\
  }\textbf {\bibinfo {volume} {100}},\ \bibinfo {pages} {107} (\bibinfo {year}
  {2018})},\ \Eprint {http://arxiv.org/abs/1711.07916} {arXiv:1711.07916
  [hep-ph]} \BibitemShut {NoStop}%
\bibitem [{\citenamefont {Holt}\ and\ \citenamefont
  {Roberts}(2010)}]{Holt:2010vj}%
  \BibitemOpen
  \bibfield  {author} {\bibinfo {author} {\bibfnamefont {R.~J.}\ \bibnamefont
  {Holt}}\ and\ \bibinfo {author} {\bibfnamefont {C.~D.}\ \bibnamefont
  {Roberts}},\ }\href {\doibase 10.1103/RevModPhys.82.2991} {\bibfield
  {journal} {\bibinfo  {journal} {Rev. Mod. Phys.}\ }\textbf {\bibinfo {volume}
  {82}},\ \bibinfo {pages} {2991} (\bibinfo {year} {2010})},\ \Eprint
  {http://arxiv.org/abs/1002.4666} {arXiv:1002.4666 [nucl-th]} \BibitemShut
  {NoStop}%
\bibitem [{\citenamefont {Mineo}\ \emph {et~al.}(2004)\citenamefont {Mineo},
  \citenamefont {Bentz}, \citenamefont {Ishii}, \citenamefont {Thomas},\ and\
  \citenamefont {Yazaki}}]{Mineo:2003vc}%
  \BibitemOpen
  \bibfield  {author} {\bibinfo {author} {\bibfnamefont {H.}~\bibnamefont
  {Mineo}}, \bibinfo {author} {\bibfnamefont {W.}~\bibnamefont {Bentz}},
  \bibinfo {author} {\bibfnamefont {N.}~\bibnamefont {Ishii}}, \bibinfo
  {author} {\bibfnamefont {A.~W.}\ \bibnamefont {Thomas}}, \ and\ \bibinfo
  {author} {\bibfnamefont {K.}~\bibnamefont {Yazaki}},\ }\href {\doibase
  10.1016/j.nuclphysa.2004.02.011} {\bibfield  {journal} {\bibinfo  {journal}
  {Nucl. Phys. A}\ }\textbf {\bibinfo {volume} {735}},\ \bibinfo {pages} {482}
  (\bibinfo {year} {2004})},\ \Eprint {http://arxiv.org/abs/nucl-th/0312097}
  {arXiv:nucl-th/0312097} \BibitemShut {NoStop}%
\bibitem [{\citenamefont {Kock}\ \emph {et~al.}(2020)\citenamefont {Kock},
  \citenamefont {Liu},\ and\ \citenamefont {Zahed}}]{Kock:2020frx}%
  \BibitemOpen
  \bibfield  {author} {\bibinfo {author} {\bibfnamefont {A.}~\bibnamefont
  {Kock}}, \bibinfo {author} {\bibfnamefont {Y.}~\bibnamefont {Liu}}, \ and\
  \bibinfo {author} {\bibfnamefont {I.}~\bibnamefont {Zahed}},\ }\href
  {\doibase 10.1103/PhysRevD.102.014039} {\bibfield  {journal} {\bibinfo
  {journal} {Phys. Rev. D}\ }\textbf {\bibinfo {volume} {102}},\ \bibinfo
  {pages} {014039} (\bibinfo {year} {2020})},\ \Eprint
  {http://arxiv.org/abs/2004.01595} {arXiv:2004.01595 [hep-ph]} \BibitemShut
  {NoStop}%
\bibitem [{\citenamefont {Wang}\ and\ \citenamefont
  {Chen}(2015)}]{Wang:2014lua}%
  \BibitemOpen
  \bibfield  {author} {\bibinfo {author} {\bibfnamefont {R.}~\bibnamefont
  {Wang}}\ and\ \bibinfo {author} {\bibfnamefont {X.}~\bibnamefont {Chen}},\
  }\href {\doibase 10.1103/PhysRevD.91.054026} {\bibfield  {journal} {\bibinfo
  {journal} {Phys. Rev. D}\ }\textbf {\bibinfo {volume} {91}},\ \bibinfo
  {pages} {054026} (\bibinfo {year} {2015})},\ \Eprint
  {http://arxiv.org/abs/1410.3598} {arXiv:1410.3598 [hep-ph]} \BibitemShut
  {NoStop}%
\bibitem [{\citenamefont {Steffens}\ and\ \citenamefont
  {Thomas}(1995)}]{Steffens:1994hf}%
  \BibitemOpen
  \bibfield  {author} {\bibinfo {author} {\bibfnamefont {F.~M.}\ \bibnamefont
  {Steffens}}\ and\ \bibinfo {author} {\bibfnamefont {A.~W.}\ \bibnamefont
  {Thomas}},\ }\href {\doibase 10.1143/PTPS.120.145} {\bibfield  {journal}
  {\bibinfo  {journal} {Prog. Theor. Phys. Suppl.}\ }\textbf {\bibinfo {volume}
  {120}},\ \bibinfo {pages} {145} (\bibinfo {year} {1995})},\ \Eprint
  {http://arxiv.org/abs/hep-ph/9509244} {arXiv:hep-ph/9509244} \BibitemShut
  {NoStop}%
\bibitem [{\citenamefont {Radyushkin}(2004)}]{Radyushkin:2004mt}%
  \BibitemOpen
  \bibfield  {author} {\bibinfo {author} {\bibfnamefont {A.}~\bibnamefont
  {Radyushkin}},\ }\href {\doibase 10.1002/andp.200410114} {\bibfield
  {journal} {\bibinfo  {journal} {Annalen Phys.}\ }\textbf {\bibinfo {volume}
  {13}},\ \bibinfo {pages} {718} (\bibinfo {year} {2004})},\ \Eprint
  {http://arxiv.org/abs/hep-ph/0410153} {arXiv:hep-ph/0410153} \BibitemShut
  {NoStop}%
\bibitem [{\citenamefont {Dokshitzer}(1977)}]{Dokshitzer:1977sg}%
  \BibitemOpen
  \bibfield  {author} {\bibinfo {author} {\bibfnamefont {Y.~L.}\ \bibnamefont
  {Dokshitzer}},\ }\href@noop {} {\bibfield  {journal} {\bibinfo  {journal}
  {Sov. Phys. JETP}\ }\textbf {\bibinfo {volume} {46}},\ \bibinfo {pages} {641}
  (\bibinfo {year} {1977})}\BibitemShut {NoStop}%
\bibitem [{\citenamefont {Gribov}\ and\ \citenamefont
  {Lipatov}(1972)}]{Gribov:1972ri}%
  \BibitemOpen
  \bibfield  {author} {\bibinfo {author} {\bibfnamefont {V.~N.}\ \bibnamefont
  {Gribov}}\ and\ \bibinfo {author} {\bibfnamefont {L.~N.}\ \bibnamefont
  {Lipatov}},\ }\href@noop {} {\bibfield  {journal} {\bibinfo  {journal} {Sov.
  J. Nucl. Phys.}\ }\textbf {\bibinfo {volume} {15}},\ \bibinfo {pages} {438}
  (\bibinfo {year} {1972})}\BibitemShut {NoStop}%
\bibitem [{\citenamefont {Altarelli}\ and\ \citenamefont
  {Parisi}(1977)}]{Altarelli:1977zs}%
  \BibitemOpen
  \bibfield  {author} {\bibinfo {author} {\bibfnamefont {G.}~\bibnamefont
  {Altarelli}}\ and\ \bibinfo {author} {\bibfnamefont {G.}~\bibnamefont
  {Parisi}},\ }\href {\doibase 10.1016/0550-3213(77)90384-4} {\bibfield
  {journal} {\bibinfo  {journal} {Nucl. Phys. B}\ }\textbf {\bibinfo {volume}
  {126}},\ \bibinfo {pages} {298} (\bibinfo {year} {1977})}\BibitemShut
  {NoStop}%
\bibitem [{\citenamefont {Nambu}\ and\ \citenamefont
  {Jona-Lasinio}(1961)}]{Nambu:1961tp}%
  \BibitemOpen
  \bibfield  {author} {\bibinfo {author} {\bibfnamefont {Y.}~\bibnamefont
  {Nambu}}\ and\ \bibinfo {author} {\bibfnamefont {G.}~\bibnamefont
  {Jona-Lasinio}},\ }\href {\doibase 10.1103/PhysRev.122.345} {\bibfield
  {journal} {\bibinfo  {journal} {Phys. Rev.}\ }\textbf {\bibinfo {volume}
  {122}},\ \bibinfo {pages} {345} (\bibinfo {year} {1961})}\BibitemShut
  {NoStop}%
\bibitem [{\citenamefont {Krein}\ \emph {et~al.}(1992)\citenamefont {Krein},
  \citenamefont {Roberts},\ and\ \citenamefont {Williams}}]{Krein:1990sf}%
  \BibitemOpen
  \bibfield  {author} {\bibinfo {author} {\bibfnamefont {G.}~\bibnamefont
  {Krein}}, \bibinfo {author} {\bibfnamefont {C.~D.}\ \bibnamefont {Roberts}},
  \ and\ \bibinfo {author} {\bibfnamefont {A.~G.}\ \bibnamefont {Williams}},\
  }\href {\doibase 10.1142/S0217751X92002544} {\bibfield  {journal} {\bibinfo
  {journal} {Int. J. Mod. Phys. A}\ }\textbf {\bibinfo {volume} {7}},\ \bibinfo
  {pages} {5607} (\bibinfo {year} {1992})}\BibitemShut {NoStop}%
\bibitem [{\citenamefont {Chang}\ \emph {et~al.}(2011)\citenamefont {Chang},
  \citenamefont {Roberts},\ and\ \citenamefont {Wilson}}]{Chang:2011zgy}%
  \BibitemOpen
  \bibfield  {author} {\bibinfo {author} {\bibfnamefont {L.}~\bibnamefont
  {Chang}}, \bibinfo {author} {\bibfnamefont {C.~D.}\ \bibnamefont {Roberts}},
  \ and\ \bibinfo {author} {\bibfnamefont {D.~J.}\ \bibnamefont {Wilson}},\
  }\href {\doibase 10.22323/1.136.0039} {\bibfield  {journal} {\bibinfo
  {journal} {PoS}\ }\textbf {\bibinfo {volume} {QCD-TNT-II2011}},\ \bibinfo
  {pages} {039} (\bibinfo {year} {2011})},\ \Eprint
  {http://arxiv.org/abs/1201.3918} {arXiv:1201.3918 [nucl-th]} \BibitemShut
  {NoStop}%
\bibitem [{\citenamefont {Roberts}(2008)}]{Roberts:2007ji}%
  \BibitemOpen
  \bibfield  {author} {\bibinfo {author} {\bibfnamefont {C.~D.}\ \bibnamefont
  {Roberts}},\ }\href {\doibase 10.1016/j.ppnp.2007.12.034} {\bibfield
  {journal} {\bibinfo  {journal} {Prog. Part. Nucl. Phys.}\ }\textbf {\bibinfo
  {volume} {61}},\ \bibinfo {pages} {50} (\bibinfo {year} {2008})},\ \Eprint
  {http://arxiv.org/abs/0712.0633} {arXiv:0712.0633 [nucl-th]} \BibitemShut
  {NoStop}%
\bibitem [{\citenamefont {Munczek}(1995)}]{Munczek:1994zz}%
  \BibitemOpen
  \bibfield  {author} {\bibinfo {author} {\bibfnamefont {H.~J.}\ \bibnamefont
  {Munczek}},\ }\href {\doibase 10.1103/PhysRevD.52.4736} {\bibfield  {journal}
  {\bibinfo  {journal} {Phys. Rev. D}\ }\textbf {\bibinfo {volume} {52}},\
  \bibinfo {pages} {4736} (\bibinfo {year} {1995})},\ \Eprint
  {http://arxiv.org/abs/hep-th/9411239} {arXiv:hep-th/9411239} \BibitemShut
  {NoStop}%
\bibitem [{\citenamefont {Aguilar}\ \emph {et~al.}(2020)\citenamefont
  {Aguilar}, \citenamefont {De~Soto}, \citenamefont {Ferreira}, \citenamefont
  {Papavassiliou}, \citenamefont {Rodr\'\i{}guez-Quintero},\ and\ \citenamefont
  {Zafeiropoulos}}]{Aguilar:2019uob}%
  \BibitemOpen
  \bibfield  {author} {\bibinfo {author} {\bibfnamefont {A.~C.}\ \bibnamefont
  {Aguilar}}, \bibinfo {author} {\bibfnamefont {F.}~\bibnamefont {De~Soto}},
  \bibinfo {author} {\bibfnamefont {M.~N.}\ \bibnamefont {Ferreira}}, \bibinfo
  {author} {\bibfnamefont {J.}~\bibnamefont {Papavassiliou}}, \bibinfo {author}
  {\bibfnamefont {J.}~\bibnamefont {Rodr\'\i{}guez-Quintero}}, \ and\ \bibinfo
  {author} {\bibfnamefont {S.}~\bibnamefont {Zafeiropoulos}},\ }\href {\doibase
  10.1140/epjc/s10052-020-7741-0} {\bibfield  {journal} {\bibinfo  {journal}
  {Eur. Phys. J. C}\ }\textbf {\bibinfo {volume} {80}},\ \bibinfo {pages} {154}
  (\bibinfo {year} {2020})},\ \Eprint {http://arxiv.org/abs/1912.12086}
  {arXiv:1912.12086 [hep-ph]} \BibitemShut {NoStop}%
\bibitem [{\citenamefont {Deur}\ \emph {et~al.}(2024)\citenamefont {Deur},
  \citenamefont {Brodsky},\ and\ \citenamefont {Roberts}}]{Deur:2023dzc}%
  \BibitemOpen
  \bibfield  {author} {\bibinfo {author} {\bibfnamefont {A.}~\bibnamefont
  {Deur}}, \bibinfo {author} {\bibfnamefont {S.~J.}\ \bibnamefont {Brodsky}}, \
  and\ \bibinfo {author} {\bibfnamefont {C.~D.}\ \bibnamefont {Roberts}},\
  }\href {\doibase 10.1016/j.ppnp.2023.104081} {\bibfield  {journal} {\bibinfo
  {journal} {Prog. Part. Nucl. Phys.}\ }\textbf {\bibinfo {volume} {134}},\
  \bibinfo {pages} {104081} (\bibinfo {year} {2024})},\ \Eprint
  {http://arxiv.org/abs/2303.00723} {arXiv:2303.00723 [hep-ph]} \BibitemShut
  {NoStop}%
\bibitem [{\citenamefont {Deur}\ \emph {et~al.}(2016)\citenamefont {Deur},
  \citenamefont {Brodsky},\ and\ \citenamefont {de~Teramond}}]{Deur:2016tte}%
  \BibitemOpen
  \bibfield  {author} {\bibinfo {author} {\bibfnamefont {A.}~\bibnamefont
  {Deur}}, \bibinfo {author} {\bibfnamefont {S.~J.}\ \bibnamefont {Brodsky}}, \
  and\ \bibinfo {author} {\bibfnamefont {G.~F.}\ \bibnamefont {de~Teramond}},\
  }\href {\doibase 10.1016/j.ppnp.2016.04.003} {\bibfield  {journal} {\bibinfo
  {journal} {Nucl. Phys.}\ }\textbf {\bibinfo {volume} {90}},\ \bibinfo {pages}
  {1} (\bibinfo {year} {2016})},\ \Eprint {http://arxiv.org/abs/1604.08082}
  {arXiv:1604.08082 [hep-ph]} \BibitemShut {NoStop}%
\bibitem [{\citenamefont {Binosi}\ \emph {et~al.}(2017)\citenamefont {Binosi},
  \citenamefont {Mezrag}, \citenamefont {Papavassiliou}, \citenamefont
  {Roberts},\ and\ \citenamefont {Rodriguez-Quintero}}]{Binosi:2016nme}%
  \BibitemOpen
  \bibfield  {author} {\bibinfo {author} {\bibfnamefont {D.}~\bibnamefont
  {Binosi}}, \bibinfo {author} {\bibfnamefont {C.}~\bibnamefont {Mezrag}},
  \bibinfo {author} {\bibfnamefont {J.}~\bibnamefont {Papavassiliou}}, \bibinfo
  {author} {\bibfnamefont {C.~D.}\ \bibnamefont {Roberts}}, \ and\ \bibinfo
  {author} {\bibfnamefont {J.}~\bibnamefont {Rodriguez-Quintero}},\ }\href
  {\doibase 10.1103/PhysRevD.96.054026} {\bibfield  {journal} {\bibinfo
  {journal} {Phys. Rev. D}\ }\textbf {\bibinfo {volume} {96}},\ \bibinfo
  {pages} {054026} (\bibinfo {year} {2017})},\ \Eprint
  {http://arxiv.org/abs/1612.04835} {arXiv:1612.04835 [nucl-th]} \BibitemShut
  {NoStop}%
\bibitem [{\citenamefont {Cui}\ \emph {et~al.}(2020{\natexlab{b}})\citenamefont
  {Cui}, \citenamefont {Zhang}, \citenamefont {Binosi}, \citenamefont
  {de~Soto}, \citenamefont {Mezrag}, \citenamefont {Papavassiliou},
  \citenamefont {Roberts}, \citenamefont {Rodr\'\i{}guez-Quintero},
  \citenamefont {Segovia},\ and\ \citenamefont {Zafeiropoulos}}]{Cui:2019dwv}%
  \BibitemOpen
  \bibfield  {author} {\bibinfo {author} {\bibfnamefont {Z.-F.}\ \bibnamefont
  {Cui}}, \bibinfo {author} {\bibfnamefont {J.-L.}\ \bibnamefont {Zhang}},
  \bibinfo {author} {\bibfnamefont {D.}~\bibnamefont {Binosi}}, \bibinfo
  {author} {\bibfnamefont {F.}~\bibnamefont {de~Soto}}, \bibinfo {author}
  {\bibfnamefont {C.}~\bibnamefont {Mezrag}}, \bibinfo {author} {\bibfnamefont
  {J.}~\bibnamefont {Papavassiliou}}, \bibinfo {author} {\bibfnamefont {C.~D.}\
  \bibnamefont {Roberts}}, \bibinfo {author} {\bibfnamefont {J.}~\bibnamefont
  {Rodr\'\i{}guez-Quintero}}, \bibinfo {author} {\bibfnamefont
  {J.}~\bibnamefont {Segovia}}, \ and\ \bibinfo {author} {\bibfnamefont
  {S.}~\bibnamefont {Zafeiropoulos}},\ }\href {\doibase
  10.1088/1674-1137/44/8/083102} {\bibfield  {journal} {\bibinfo  {journal}
  {Chin. Phys. C}\ }\textbf {\bibinfo {volume} {44}},\ \bibinfo {pages}
  {083102} (\bibinfo {year} {2020}{\natexlab{b}})},\ \Eprint
  {http://arxiv.org/abs/1912.08232} {arXiv:1912.08232 [hep-ph]} \BibitemShut
  {NoStop}%
\bibitem [{\citenamefont {Gribov}\ \emph {et~al.}(1983)\citenamefont {Gribov},
  \citenamefont {Levin},\ and\ \citenamefont {Ryskin}}]{Gribov:1983ivg}%
  \BibitemOpen
  \bibfield  {author} {\bibinfo {author} {\bibfnamefont {L.~V.}\ \bibnamefont
  {Gribov}}, \bibinfo {author} {\bibfnamefont {E.~M.}\ \bibnamefont {Levin}}, \
  and\ \bibinfo {author} {\bibfnamefont {M.~G.}\ \bibnamefont {Ryskin}},\
  }\href {\doibase 10.1016/0370-1573(83)90022-4} {\bibfield  {journal}
  {\bibinfo  {journal} {Phys. Rept.}\ }\textbf {\bibinfo {volume} {100}},\
  \bibinfo {pages} {1} (\bibinfo {year} {1983})}\BibitemShut {NoStop}%
\bibitem [{\citenamefont {Mueller}\ and\ \citenamefont
  {Qiu}(1986)}]{Mueller:1985wy}%
  \BibitemOpen
  \bibfield  {author} {\bibinfo {author} {\bibfnamefont {A.~H.}\ \bibnamefont
  {Mueller}}\ and\ \bibinfo {author} {\bibfnamefont {J.-w.}\ \bibnamefont
  {Qiu}},\ }\href {\doibase 10.1016/0550-3213(86)90164-1} {\bibfield  {journal}
  {\bibinfo  {journal} {Nucl. Phys. B}\ }\textbf {\bibinfo {volume} {268}},\
  \bibinfo {pages} {427} (\bibinfo {year} {1986})}\BibitemShut {NoStop}%
\bibitem [{\citenamefont {Zhu}(1999)}]{Zhu:1998hg}%
  \BibitemOpen
  \bibfield  {author} {\bibinfo {author} {\bibfnamefont {W.}~\bibnamefont
  {Zhu}},\ }\href {\doibase 10.1016/S0550-3213(99)00237-0} {\bibfield
  {journal} {\bibinfo  {journal} {Nucl. Phys. B}\ }\textbf {\bibinfo {volume}
  {551}},\ \bibinfo {pages} {245} (\bibinfo {year} {1999})},\ \Eprint
  {http://arxiv.org/abs/hep-ph/9809391} {arXiv:hep-ph/9809391} \BibitemShut
  {NoStop}%
\bibitem [{\citenamefont {Zhu}\ and\ \citenamefont {Ruan}(1999)}]{Zhu:1999ht}%
  \BibitemOpen
  \bibfield  {author} {\bibinfo {author} {\bibfnamefont {W.}~\bibnamefont
  {Zhu}}\ and\ \bibinfo {author} {\bibfnamefont {J.-h.}\ \bibnamefont {Ruan}},\
  }\href {\doibase 10.1016/S0550-3213(99)00461-7} {\bibfield  {journal}
  {\bibinfo  {journal} {Nucl. Phys. B}\ }\textbf {\bibinfo {volume} {559}},\
  \bibinfo {pages} {378} (\bibinfo {year} {1999})},\ \Eprint
  {http://arxiv.org/abs/hep-ph/9907330} {arXiv:hep-ph/9907330} \BibitemShut
  {NoStop}%
\bibitem [{\citenamefont {Zhu}\ and\ \citenamefont {Shen}(2005)}]{Zhu:2004xj}%
  \BibitemOpen
  \bibfield  {author} {\bibinfo {author} {\bibfnamefont {W.}~\bibnamefont
  {Zhu}}\ and\ \bibinfo {author} {\bibfnamefont {Z.-q.}\ \bibnamefont {Shen}},\
  }\href@noop {} {\bibfield  {journal} {\bibinfo  {journal} {HEP. \& NP. Vol.}\
  }\textbf {\bibinfo {volume} {2}},\ \bibinfo {pages} {,109} (\bibinfo {year}
  {2005})},\ \Eprint {http://arxiv.org/abs/hep-ph/0406213}
  {arXiv:hep-ph/0406213} \BibitemShut {NoStop}%
\bibitem [{\citenamefont {Gell-Mann}(1964)}]{Gell-Mann:1964ewy}%
  \BibitemOpen
  \bibfield  {author} {\bibinfo {author} {\bibfnamefont {M.}~\bibnamefont
  {Gell-Mann}},\ }\href {\doibase 10.1016/S0031-9163(64)92001-3} {\bibfield
  {journal} {\bibinfo  {journal} {Phys. Lett.}\ }\textbf {\bibinfo {volume}
  {8}},\ \bibinfo {pages} {214} (\bibinfo {year} {1964})}\BibitemShut {NoStop}%
\bibitem [{\citenamefont {Zweig}(1964)}]{Zweig:1964ruk}%
  \BibitemOpen
  \bibfield  {author} {\bibinfo {author} {\bibfnamefont {G.}~\bibnamefont
  {Zweig}},\ }\href {\doibase 10.17181/CERN-TH-401} {\  (\bibinfo {year}
  {1964}),\ 10.17181/CERN-TH-401}\BibitemShut {NoStop}%
\bibitem [{\citenamefont {Giannini}(1991)}]{Giannini:1990pc}%
  \BibitemOpen
  \bibfield  {author} {\bibinfo {author} {\bibfnamefont {M.~M.}\ \bibnamefont
  {Giannini}},\ }\href {\doibase 10.1088/0034-4885/54/3/003} {\bibfield
  {journal} {\bibinfo  {journal} {Rept. Prog. Phys.}\ }\textbf {\bibinfo
  {volume} {54}},\ \bibinfo {pages} {453} (\bibinfo {year} {1991})}\BibitemShut
  {NoStop}%
\bibitem [{\citenamefont {Eichmann}\ \emph {et~al.}(2016)\citenamefont
  {Eichmann}, \citenamefont {Sanchis-Alepuz}, \citenamefont {Williams},
  \citenamefont {Alkofer},\ and\ \citenamefont {Fischer}}]{Eichmann:2016yit}%
  \BibitemOpen
  \bibfield  {author} {\bibinfo {author} {\bibfnamefont {G.}~\bibnamefont
  {Eichmann}}, \bibinfo {author} {\bibfnamefont {H.}~\bibnamefont
  {Sanchis-Alepuz}}, \bibinfo {author} {\bibfnamefont {R.}~\bibnamefont
  {Williams}}, \bibinfo {author} {\bibfnamefont {R.}~\bibnamefont {Alkofer}}, \
  and\ \bibinfo {author} {\bibfnamefont {C.~S.}\ \bibnamefont {Fischer}},\
  }\href {\doibase 10.1016/j.ppnp.2016.07.001} {\bibfield  {journal} {\bibinfo
  {journal} {Prog. Part. Nucl. Phys.}\ }\textbf {\bibinfo {volume} {91}},\
  \bibinfo {pages} {1} (\bibinfo {year} {2016})},\ \Eprint
  {http://arxiv.org/abs/1606.09602} {arXiv:1606.09602 [hep-ph]} \BibitemShut
  {NoStop}%
\bibitem [{\citenamefont {Altarelli}\ \emph {et~al.}(1974)\citenamefont
  {Altarelli}, \citenamefont {Cabibbo}, \citenamefont {Maiani},\ and\
  \citenamefont {Petronzio}}]{Altarelli:1973ff}%
  \BibitemOpen
  \bibfield  {author} {\bibinfo {author} {\bibfnamefont {G.}~\bibnamefont
  {Altarelli}}, \bibinfo {author} {\bibfnamefont {N.}~\bibnamefont {Cabibbo}},
  \bibinfo {author} {\bibfnamefont {L.}~\bibnamefont {Maiani}}, \ and\ \bibinfo
  {author} {\bibfnamefont {R.}~\bibnamefont {Petronzio}},\ }\href {\doibase
  10.1016/0550-3213(74)90452-0} {\bibfield  {journal} {\bibinfo  {journal}
  {Nucl. Phys. B}\ }\textbf {\bibinfo {volume} {69}},\ \bibinfo {pages} {531}
  (\bibinfo {year} {1974})}\BibitemShut {NoStop}%
\bibitem [{\citenamefont {Parisi}\ and\ \citenamefont
  {Petronzio}(1976)}]{Parisi:1976fz}%
  \BibitemOpen
  \bibfield  {author} {\bibinfo {author} {\bibfnamefont {G.}~\bibnamefont
  {Parisi}}\ and\ \bibinfo {author} {\bibfnamefont {R.}~\bibnamefont
  {Petronzio}},\ }\href {\doibase 10.1016/0370-2693(76)90088-5} {\bibfield
  {journal} {\bibinfo  {journal} {Phys. Lett. B}\ }\textbf {\bibinfo {volume}
  {62}},\ \bibinfo {pages} {331} (\bibinfo {year} {1976})}\BibitemShut
  {NoStop}%
\bibitem [{\citenamefont {Gluck}\ and\ \citenamefont
  {Reya}(1977)}]{Gluck:1977ah}%
  \BibitemOpen
  \bibfield  {author} {\bibinfo {author} {\bibfnamefont {M.}~\bibnamefont
  {Gluck}}\ and\ \bibinfo {author} {\bibfnamefont {E.}~\bibnamefont {Reya}},\
  }\href {\doibase 10.1016/0550-3213(77)90393-5} {\bibfield  {journal}
  {\bibinfo  {journal} {Nucl. Phys. B}\ }\textbf {\bibinfo {volume} {130}},\
  \bibinfo {pages} {76} (\bibinfo {year} {1977})}\BibitemShut {NoStop}%
\bibitem [{\citenamefont {Vainshtein}\ \emph {et~al.}(1976)\citenamefont
  {Vainshtein}, \citenamefont {Zakharov}, \citenamefont {Novikov},\ and\
  \citenamefont {Shifman}}]{Vainshtein:1976kd}%
  \BibitemOpen
  \bibfield  {author} {\bibinfo {author} {\bibfnamefont {A.~I.}\ \bibnamefont
  {Vainshtein}}, \bibinfo {author} {\bibfnamefont {V.~I.}\ \bibnamefont
  {Zakharov}}, \bibinfo {author} {\bibfnamefont {V.~A.}\ \bibnamefont
  {Novikov}}, \ and\ \bibinfo {author} {\bibfnamefont {M.~A.}\ \bibnamefont
  {Shifman}},\ }\href@noop {} {\bibfield  {journal} {\bibinfo  {journal} {JETP
  Lett.}\ }\textbf {\bibinfo {volume} {24}},\ \bibinfo {pages} {341} (\bibinfo
  {year} {1976})}\BibitemShut {NoStop}%
\bibitem [{\citenamefont {Collins}\ and\ \citenamefont
  {Qiu}(1989)}]{Collins:1988wj}%
  \BibitemOpen
  \bibfield  {author} {\bibinfo {author} {\bibfnamefont {J.~C.}\ \bibnamefont
  {Collins}}\ and\ \bibinfo {author} {\bibfnamefont {J.-w.}\ \bibnamefont
  {Qiu}},\ }\href {\doibase 10.1103/PhysRevD.39.1398} {\bibfield  {journal}
  {\bibinfo  {journal} {Phys. Rev. D}\ }\textbf {\bibinfo {volume} {39}},\
  \bibinfo {pages} {1398} (\bibinfo {year} {1989})}\BibitemShut {NoStop}%
\bibitem [{\citenamefont {Gl\"uck}\ \emph {et~al.}(1998)\citenamefont
  {Gl\"uck}, \citenamefont {Reya},\ and\ \citenamefont {Vogt}}]{Gluck:1998xa}%
  \BibitemOpen
  \bibfield  {author} {\bibinfo {author} {\bibfnamefont {M.}~\bibnamefont
  {Gl\"uck}}, \bibinfo {author} {\bibfnamefont {E.}~\bibnamefont {Reya}}, \
  and\ \bibinfo {author} {\bibfnamefont {A.}~\bibnamefont {Vogt}},\ }\href
  {\doibase 10.1007/s100520050289} {\bibfield  {journal} {\bibinfo  {journal}
  {Eur. Phys. J. C}\ }\textbf {\bibinfo {volume} {5}},\ \bibinfo {pages} {461}
  (\bibinfo {year} {1998})},\ \Eprint {http://arxiv.org/abs/hep-ph/9806404}
  {arXiv:hep-ph/9806404} \BibitemShut {NoStop}%
\bibitem [{\citenamefont {Wang}\ and\ \citenamefont
  {Chen}(2017)}]{Wang:2016sfq}%
  \BibitemOpen
  \bibfield  {author} {\bibinfo {author} {\bibfnamefont {R.}~\bibnamefont
  {Wang}}\ and\ \bibinfo {author} {\bibfnamefont {X.}~\bibnamefont {Chen}},\
  }\href {\doibase 10.1088/1674-1137/41/5/053103} {\bibfield  {journal}
  {\bibinfo  {journal} {Chin. Phys. C}\ }\textbf {\bibinfo {volume} {41}},\
  \bibinfo {pages} {053103} (\bibinfo {year} {2017})},\ \Eprint
  {http://arxiv.org/abs/1609.01831} {arXiv:1609.01831 [hep-ph]} \BibitemShut
  {NoStop}%
\bibitem [{\citenamefont {Zhu}\ and\ \citenamefont {Wang}(2019)}]{Zhu:2019ydy}%
  \BibitemOpen
  \bibfield  {author} {\bibinfo {author} {\bibfnamefont {W.}~\bibnamefont
  {Zhu}}\ and\ \bibinfo {author} {\bibfnamefont {F.}~\bibnamefont {Wang}},\
  }\href@noop {} {\  (\bibinfo {year} {2019})},\ \Eprint
  {http://arxiv.org/abs/1912.09785} {arXiv:1912.09785 [hep-ph]} \BibitemShut
  {NoStop}%
\bibitem [{\citenamefont {Froissart}(1961)}]{Froissart:1961ux}%
  \BibitemOpen
  \bibfield  {author} {\bibinfo {author} {\bibfnamefont {M.}~\bibnamefont
  {Froissart}},\ }\href {\doibase 10.1103/PhysRev.123.1053} {\bibfield
  {journal} {\bibinfo  {journal} {Phys. Rev.}\ }\textbf {\bibinfo {volume}
  {123}},\ \bibinfo {pages} {1053} (\bibinfo {year} {1961})}\BibitemShut
  {NoStop}%
\bibitem [{\citenamefont {Martin}(1963)}]{Martin:1962rt}%
  \BibitemOpen
  \bibfield  {author} {\bibinfo {author} {\bibfnamefont {A.}~\bibnamefont
  {Martin}},\ }\href {\doibase 10.1103/PhysRev.129.1432} {\bibfield  {journal}
  {\bibinfo  {journal} {Phys. Rev.}\ }\textbf {\bibinfo {volume} {129}},\
  \bibinfo {pages} {1432} (\bibinfo {year} {1963})}\BibitemShut {NoStop}%
\bibitem [{\citenamefont {Chen}\ \emph {et~al.}(2016)\citenamefont {Chen},
  \citenamefont {Ruan}, \citenamefont {Wang}, \citenamefont {Zhang},\ and\
  \citenamefont {Zhu}}]{Chen:2014nba}%
  \BibitemOpen
  \bibfield  {author} {\bibinfo {author} {\bibfnamefont {X.}~\bibnamefont
  {Chen}}, \bibinfo {author} {\bibfnamefont {J.}~\bibnamefont {Ruan}}, \bibinfo
  {author} {\bibfnamefont {R.}~\bibnamefont {Wang}}, \bibinfo {author}
  {\bibfnamefont {P.}~\bibnamefont {Zhang}}, \ and\ \bibinfo {author}
  {\bibfnamefont {W.}~\bibnamefont {Zhu}},\ }\href {\doibase
  10.1140/epjp/i2016-16006-x} {\bibfield  {journal} {\bibinfo  {journal} {Eur.
  Phys. J. Plus}\ }\textbf {\bibinfo {volume} {131}},\ \bibinfo {pages} {6}
  (\bibinfo {year} {2016})},\ \Eprint {http://arxiv.org/abs/1404.0759}
  {arXiv:1404.0759 [hep-ph]} \BibitemShut {NoStop}%
\bibitem [{\citenamefont {Whitlow}\ \emph {et~al.}(1992)\citenamefont
  {Whitlow}, \citenamefont {Riordan}, \citenamefont {Dasu}, \citenamefont
  {Rock},\ and\ \citenamefont {Bodek}}]{Whitlow:1991uw}%
  \BibitemOpen
  \bibfield  {author} {\bibinfo {author} {\bibfnamefont {L.~W.}\ \bibnamefont
  {Whitlow}}, \bibinfo {author} {\bibfnamefont {E.~M.}\ \bibnamefont
  {Riordan}}, \bibinfo {author} {\bibfnamefont {S.}~\bibnamefont {Dasu}},
  \bibinfo {author} {\bibfnamefont {S.}~\bibnamefont {Rock}}, \ and\ \bibinfo
  {author} {\bibfnamefont {A.}~\bibnamefont {Bodek}},\ }\href {\doibase
  10.1016/0370-2693(92)90672-Q} {\bibfield  {journal} {\bibinfo  {journal}
  {Phys. Lett. B}\ }\textbf {\bibinfo {volume} {282}},\ \bibinfo {pages} {475}
  (\bibinfo {year} {1992})}\BibitemShut {NoStop}%
\bibitem [{\citenamefont {Aaron}\ \emph {et~al.}(2010)\citenamefont {Aaron}
  \emph {et~al.}}]{H1:2009pze}%
  \BibitemOpen
  \bibfield  {author} {\bibinfo {author} {\bibfnamefont {F.~D.}\ \bibnamefont
  {Aaron}} \emph {et~al.} (\bibinfo {collaboration} {H1, ZEUS}),\ }\href
  {\doibase 10.1007/JHEP01(2010)109} {\bibfield  {journal} {\bibinfo  {journal}
  {JHEP}\ }\textbf {\bibinfo {volume} {01}},\ \bibinfo {pages} {109} (\bibinfo
  {year} {2010})},\ \Eprint {http://arxiv.org/abs/0911.0884} {arXiv:0911.0884
  [hep-ex]} \BibitemShut {NoStop}%
\bibitem [{\citenamefont {Brodsky}\ \emph {et~al.}(1980)\citenamefont
  {Brodsky}, \citenamefont {Hoyer}, \citenamefont {Peterson},\ and\
  \citenamefont {Sakai}}]{Brodsky:1980pb}%
  \BibitemOpen
  \bibfield  {author} {\bibinfo {author} {\bibfnamefont {S.~J.}\ \bibnamefont
  {Brodsky}}, \bibinfo {author} {\bibfnamefont {P.}~\bibnamefont {Hoyer}},
  \bibinfo {author} {\bibfnamefont {C.}~\bibnamefont {Peterson}}, \ and\
  \bibinfo {author} {\bibfnamefont {N.}~\bibnamefont {Sakai}},\ }\href
  {\doibase 10.1016/0370-2693(80)90364-0} {\bibfield  {journal} {\bibinfo
  {journal} {Phys. Lett. B}\ }\textbf {\bibinfo {volume} {93}},\ \bibinfo
  {pages} {451} (\bibinfo {year} {1980})}\BibitemShut {NoStop}%
\bibitem [{\citenamefont {Zou}\ and\ \citenamefont {Riska}(2005)}]{Zou:2005xy}%
  \BibitemOpen
  \bibfield  {author} {\bibinfo {author} {\bibfnamefont {B.~S.}\ \bibnamefont
  {Zou}}\ and\ \bibinfo {author} {\bibfnamefont {D.~O.}\ \bibnamefont
  {Riska}},\ }\href {\doibase 10.1103/PhysRevLett.95.072001} {\bibfield
  {journal} {\bibinfo  {journal} {Phys. Rev. Lett.}\ }\textbf {\bibinfo
  {volume} {95}},\ \bibinfo {pages} {072001} (\bibinfo {year} {2005})},\
  \Eprint {http://arxiv.org/abs/hep-ph/0502225} {arXiv:hep-ph/0502225}
  \BibitemShut {NoStop}%
\bibitem [{\citenamefont {Zou}(2010)}]{Zou:2010tc}%
  \BibitemOpen
  \bibfield  {author} {\bibinfo {author} {\bibfnamefont {B.-S.}\ \bibnamefont
  {Zou}},\ }\href {\doibase 10.1016/j.nuclphysa.2010.01.194} {\bibfield
  {journal} {\bibinfo  {journal} {Nucl. Phys. A}\ }\textbf {\bibinfo {volume}
  {835}},\ \bibinfo {pages} {199} (\bibinfo {year} {2010})},\ \Eprint
  {http://arxiv.org/abs/1001.1084} {arXiv:1001.1084 [nucl-th]} \BibitemShut
  {NoStop}%
\bibitem [{\citenamefont {Chang}\ and\ \citenamefont
  {Peng}(2011)}]{Chang:2011vx}%
  \BibitemOpen
  \bibfield  {author} {\bibinfo {author} {\bibfnamefont {W.-C.}\ \bibnamefont
  {Chang}}\ and\ \bibinfo {author} {\bibfnamefont {J.-C.}\ \bibnamefont
  {Peng}},\ }\href {\doibase 10.1103/PhysRevLett.106.252002} {\bibfield
  {journal} {\bibinfo  {journal} {Phys. Rev. Lett.}\ }\textbf {\bibinfo
  {volume} {106}},\ \bibinfo {pages} {252002} (\bibinfo {year} {2011})},\
  \Eprint {http://arxiv.org/abs/1102.5631} {arXiv:1102.5631 [hep-ph]}
  \BibitemShut {NoStop}%
\bibitem [{\citenamefont {Liu}\ \emph {et~al.}(2012)\citenamefont {Liu},
  \citenamefont {Chang}, \citenamefont {Cheng},\ and\ \citenamefont
  {Peng}}]{Liu:2012ch}%
  \BibitemOpen
  \bibfield  {author} {\bibinfo {author} {\bibfnamefont {K.-F.}\ \bibnamefont
  {Liu}}, \bibinfo {author} {\bibfnamefont {W.-C.}\ \bibnamefont {Chang}},
  \bibinfo {author} {\bibfnamefont {H.-Y.}\ \bibnamefont {Cheng}}, \ and\
  \bibinfo {author} {\bibfnamefont {J.-C.}\ \bibnamefont {Peng}},\ }\href
  {\doibase 10.1103/PhysRevLett.109.252002} {\bibfield  {journal} {\bibinfo
  {journal} {Phys. Rev. Lett.}\ }\textbf {\bibinfo {volume} {109}},\ \bibinfo
  {pages} {252002} (\bibinfo {year} {2012})},\ \Eprint
  {http://arxiv.org/abs/1206.4339} {arXiv:1206.4339 [hep-ph]} \BibitemShut
  {NoStop}%
\bibitem [{\citenamefont {Signal}\ \emph {et~al.}(1991)\citenamefont {Signal},
  \citenamefont {Schreiber},\ and\ \citenamefont {Thomas}}]{Signal:1991ug}%
  \BibitemOpen
  \bibfield  {author} {\bibinfo {author} {\bibfnamefont {A.~I.}\ \bibnamefont
  {Signal}}, \bibinfo {author} {\bibfnamefont {A.~W.}\ \bibnamefont
  {Schreiber}}, \ and\ \bibinfo {author} {\bibfnamefont {A.~W.}\ \bibnamefont
  {Thomas}},\ }\href {\doibase 10.1142/S0217732391000233} {\bibfield  {journal}
  {\bibinfo  {journal} {Mod. Phys. Lett. A}\ }\textbf {\bibinfo {volume} {6}},\
  \bibinfo {pages} {271} (\bibinfo {year} {1991})}\BibitemShut {NoStop}%
\bibitem [{\citenamefont {Melnitchouk}\ \emph {et~al.}(1998)\citenamefont
  {Melnitchouk}, \citenamefont {Speth},\ and\ \citenamefont
  {Thomas}}]{Melnitchouk:1998rv}%
  \BibitemOpen
  \bibfield  {author} {\bibinfo {author} {\bibfnamefont {W.}~\bibnamefont
  {Melnitchouk}}, \bibinfo {author} {\bibfnamefont {J.}~\bibnamefont {Speth}},
  \ and\ \bibinfo {author} {\bibfnamefont {A.~W.}\ \bibnamefont {Thomas}},\
  }\href {\doibase 10.1103/PhysRevD.59.014033} {\bibfield  {journal} {\bibinfo
  {journal} {Phys. Rev. D}\ }\textbf {\bibinfo {volume} {59}},\ \bibinfo
  {pages} {014033} (\bibinfo {year} {1998})},\ \Eprint
  {http://arxiv.org/abs/hep-ph/9806255} {arXiv:hep-ph/9806255} \BibitemShut
  {NoStop}%
\bibitem [{\citenamefont {Nikolaev}\ \emph {et~al.}(1999)\citenamefont
  {Nikolaev}, \citenamefont {Schafer}, \citenamefont {Szczurek},\ and\
  \citenamefont {Speth}}]{Nikolaev:1998se}%
  \BibitemOpen
  \bibfield  {author} {\bibinfo {author} {\bibfnamefont {N.~N.}\ \bibnamefont
  {Nikolaev}}, \bibinfo {author} {\bibfnamefont {W.}~\bibnamefont {Schafer}},
  \bibinfo {author} {\bibfnamefont {A.}~\bibnamefont {Szczurek}}, \ and\
  \bibinfo {author} {\bibfnamefont {J.}~\bibnamefont {Speth}},\ }\href
  {\doibase 10.1103/PhysRevD.60.014004} {\bibfield  {journal} {\bibinfo
  {journal} {Phys. Rev. D}\ }\textbf {\bibinfo {volume} {60}},\ \bibinfo
  {pages} {014004} (\bibinfo {year} {1999})},\ \Eprint
  {http://arxiv.org/abs/hep-ph/9812266} {arXiv:hep-ph/9812266} \BibitemShut
  {NoStop}%
\end{thebibliography}%

\end{document}